\documentclass[12pt]{article}
\usepackage{amsfonts,amssymb,amsmath}
\usepackage{mathtools}
\usepackage{graphicx}
\usepackage{wrapfig}
\usepackage{caption}
\usepackage{comment}
\usepackage[colorlinks=true,citecolor=blue,linkcolor=black]{hyperref}
\usepackage{url}
\usepackage{cite}
\usepackage{xspace}
\usepackage[nottoc]{tocbibind}
\usepackage{indentfirst}

\usepackage{scalerel,adjustbox}

\makeatletter
\newcommand*\wt[1]{\mathpalette\wthelper{#1}}
\newcommand*\wthelper[2]{%
        \hbox{\dimen@\accentfontxheight#1%
                \accentfontxheight#11.15\dimen@
                $\m@th#1\widetilde{#2}$%
                \accentfontxheight#1\dimen@
        }%
}

\newcommand*\accentfontxheight[1]{%
        \fontdimen5\ifx#1\displaystyle
                \textfont
        \else\ifx#1\textstyle
                \textfont
        \else\ifx#1\scriptstyle
                \scriptfont
        \else
                \scriptscriptfont
        \fi\fi\fi3
}
\makeatother

\newcommand*{\Tr}{\ensuremath{\mathbf{tr}}}

\renewcommand*{\mod}{\ensuremath{\mathrel{\mathrm{mod}}}}

\begin{document}

\title{\vspace{-.5cm}\Large\bf An Explicit Categorical Construction of \\ Instanton Density in Lattice Yang-Mills Theory}
\author{ \large Peng Zhang and Jing-Yuan Chen \\[.2cm]
{\small\em Institute for Advanced Study, Tsinghua University, Beijing, 100084, China}}

\date{}

\maketitle

\begin{abstract}
Since the inception of lattice QCD, a natural definition for the Yang-Mills instanton on lattice has been long sought for. In a recent work \cite{Chen:2024ddr}, one of authors showed the natural solution has to be organized in terms of bundle gerbes in higher homotopy theory / higher category theory, and introduced the principles for such a categorical construction. To pave the way towards actual numerical implementation in the near future, nonetheless, an explicit construction is necessary. In this paper we provide such an explicit construction for $SU(2)$ gauge theory, with technical aspects inspired by L\"{u}scher's 1982 geometrical construction \cite{Luscher:1981zq}. We will see how the latter is in a suitable sense a saddle point approximation to the full categorical construction. The generalization to $SU(N)$ will be discussed. The construction also allows for a natural definition of lattice Chern-Simons-Yang-Mills theory in three spacetime dimensions.\end{abstract}

\tableofcontents


\section{Introduction}
\label{s_intro}

An important approach to understand quantum chromodynamics (QCD) is lattice QCD, at both the conceptual and the practical level. At the conceptual level, the lattice provides a non-perturbative UV regularization for QCD \cite{Wilson:1974sk}. At the practical level, Monte-Carlo numerical computation can be performed in order to study the otherwise hard to extract strongly coupled physics of the theory \cite{Creutz:1979dw, Wilson:1979wp}. Tremendous success has been achieved since the inception of this idea half a century ago.

Nevertheless, since the invention of lattice QCD, there has been a long standing problem, that the \emph{instanton configuration} \cite{Belavin:1975fg}, a topological configuration of the Yang-Mills gauge field that has important physical consequences (see e.g. \cite{Schafer:1996wv} for a review), does not admit a natural definition on the lattice. More precisely, while the physical effects of the instanton configurations must have already been somehow included in the lattice Yang-Mills path integral, there is no natural way to sharply define the instanton itself in terms of the lattice Yang-Mills degrees of freedom, thus making it hard to adequately extract the associated physical effects, let alone to effectively control the instanton fugacity on the lattice. Started with \cite{Luscher:1981zq}, there has been a lot of efforts on workaround solutions, see e.g. \cite{Alexandrou:2017hqw} for a review.

In a recent work \cite{Chen:2024ddr}, one of the authors showed that, to naturally define instanton on the lattice, the implementation of Yang-Mills gauge field on the lattice must be refined beyond Wilson's original definition \cite{Wilson:1974sk}, such that in addition to the original Wilson line degrees of freedom on the lattice links, there must also be suitable degrees of freedom on the lattice plaquettes, cubes and hypercubes to be sampled in the lattice path integral.
\footnote{As we will see, this is conceptually similar to, and actually inspired by, the Villain model \cite{Berezinsky:1970fr,Villain:1974ir}.}
Importantly, the introduction of these new degrees of freedom is not \emph{ad hoc}, but deeply rooted in mathematics: The degrees of freedom altogether form a higher category structure (a suitable weak 4-group) that implements the second Chern class on the lattice; moreover, it is mathematically impossible to reduce this structure anyhow such that it becomes describable in terms of the familiar notions of Lie groups and/or fibre bundles. 
\footnote{This might be why, previously, a natural definition for lattice Yang-Mills instanton has been hard to find. This aspect deviates from the Villain model, as the latter can be described in terms of principal bundle.}
Therefore, the language of higher category theory really is necessary here, rather than just being fancy.

While ``higher category'' might sound highly abstract, the idea is physically very intuitive: Given a lattice gauge field configuration on the links, the ways of interpolating it into a continuum gauge field configuration is obviously non-unique; even if we ignore some UV details, at the topological level there are still topologically distinct possibilities. Indeed, the new degrees of freedom are therefore introduced to adequately sample these different possibilities. The principles of how to achieve such a construction has been explained in \cite{Chen:2024ddr}, along with the mathematical framework underneath. In order to facilitate actual numerical implementation in the near future, however, it is important to have a more explicit construction. This is the goal of this paper. 

As we will see, some technical parts in our construction are inspired by L\"{u}scher's geometrical construction \cite{Luscher:1981zq}, but in a conceptually different way. Through this, we will see in retrospect how L\"{u}scher's construction appears to be a saddle point approximation to the fully-fledged categorical construction, with all those new degrees of freedom that we shall introduce taking values that maximize the path integral weight.

In addition to defining instanton density in 4d spacetime lattice, the refined lattice Yang-Mills degrees of freedom also allows Chern-Simons-Yang-Mills theory to be naturally defined on 3d spacetime lattice. While such a lattice theory cannot be studied by Monte-Carlo due to the complex Chern-Simons (CS) phase, nor can it be solved analytically, it is conceptually important to have such a lattice definition, and it will be interesting to analyze the properties of the theory in future works.

\

In this paper we will only describe the intuitive explicit construction, while directing any mathematical formality to \cite{Chen:2024ddr}. This paper is organized as the following. In Section \ref{s_path_int}, we will review the problem, introduce the basic idea of the solution, and setup the lattice path integral for $SU(2)$ gauge field. In Section \ref{s_CS}, we will describe in details the construction of the CS phase saddle around the lattice hypercube, which is the key technical part of the construction. In Section \ref{s_SUN}, we will discuss the generalization to $SU(N)$ gauge field. In Section \ref{s_remarks} we will give some further remarks.

\section{The Lattice Path Integral for $SU(2)$}
\label{s_path_int}

We begin by reviewing the origin of the instanton definition problem. In the continuum, the instanton number (also called topological charge) $Q$, as an integral of the instanton density $\mathcal{Q}$ over the spacetime, is 
\begin{equation}
    Q=\int_{4d} \mathcal{Q}=\frac{1}{2(2\pi)^2}\int_{4d}\Tr(F\wedge F),\quad F=\mathrm{d}A-iA\wedge A \ ,
    \label{eqn:q_def}
\end{equation}
and can be shown to be an integer. In Wilson's lattice gauge theory \cite{Wilson:1974sk} (in this paper it suffices to study pure gauge theory), however, we have only the Lie group valued holonomy around each plaquette, but not a Lie algebra valued curvature $F$:
\begin{equation}
    Z=\left[\prod_{l'} \int_G\mathrm{d}u_{l'}\right] W^{(0)}[u], \ \ \ \ \ W^{(0)}[u]=\prod_p W_2^{(0)}(\Tr \, u_p+c.c.)
    \label{eqn:wilson_weight}
\end{equation}
where the group valued dynamical field $u_l\in G$ is the gauge field's Wilson line across the link $l$, from which we can build $u_p$, the Wilson loop holonomy around plaquette $p$. $W_2^{(0)}(\Tr \, u_p+c.c.)$ is the weight on the plaquette (the subscript $_2$ means the plaquette is 2d); it should be a positive and decreasing function, but otherwise its detailed form does not have to be specified and shall in practice be optimized numerically. While a Lie algebra element can be obtained from a Lie group element by taking the logarithm, such as $F_{p}\overset{\text{naive}}{=}-i\ln u_p$, the logarithm is not continuous and not single-valued.
\footnote{Some other desired properties are also lacked, because $e^{X+Y}\neq e^X e^Y$ when the group is non-abelian. But the crucial problem is the discontinuity.} 
Hence it is not evident that the expression (\ref{eqn:q_def}) has a good counterpart on the lattice.

In fact, a more general consideration shows there \emph{cannot possibly be} a natural counterpart. In Wilson's lattice gauge theory  \cite{Wilson:1974sk}, the configuration space consists of a group element $u_l\in G$ on each link $l$, so the total configuration space is $G\times \cdots \times G=G^{|\text{number of links}|}$.
\footnote{As emphasized by Wilson \cite{Wilson:1974sk}, gauge redundancy does not (and must not) require any treatment on the lattice. To the path integral, gauge redundancy only contributes a \emph{local product} of finite constant factors. For the observables, Elitzur's theorem \cite{Elitzur:1975im} ensure the gauge non-invariant observables essentially vanish anyways.}
If there could be \emph{any} lattice counterpart of Eq.(\ref{eqn:q_def}), where the gauge field configuration determines the instanton number, then we will be looking at a map from the lattice configuration space to the integers:
\begin{equation}
   Q_{\text{lattice, naive}}: G^{|\text{number of links}|} \ \longrightarrow \ \mathbb{Z} \ .
   \label{eqn:origin}
\end{equation}
The key observation is, for connected Lie groups such as $G=SU(N)$, the left-hand-side is a connected space (that is to say, any two gauge field configurations in Wilson's lattice gauge theory can be continuously deformed to one another), while the right-hand-side is discrete. Thus, if we want the instanton number to be non-constant, the map must be discontinuous---which is fundamentally/mathematically unnatural, and will lead to various problems in practice. Such discontinuity is indeed present in the current workaround solutions \cite{Alexandrou:2017hqw}.
\footnote{There are associated extra problems in the current solutions \cite{Chen:2024ddr}: The geometrical construction \cite{Luscher:1981zq,Phillips:1986qd} discards a large portion of the configuration space (see Appendix \ref{app:q_density} for relevant discussions). The flow methods will change the local instanton density so that only ``large instantons'' remain (at least intuitively, since the density really is not well-defined). The fermion index methods, among other things, require a spin structure (fermion boundary condition) on the lattice, which should conceptually be irrelevant in defining a Yang-Mills theory.}
This is the simple, general proof that a natural definition for instanton is impossible in Wilson's lattice gauge theory.

When $G=U(1)$, there is a well-known approach to naturally to define topological configurations on the lattice. This is the method of Villainization \cite{Berezinsky:1970fr, Villain:1974ir}, first developed by Berezinsky in the XY model, i.e. $S^1$ lattice non-linear sigma model, to naturally define and study winding and vortices \cite{Berezinsky:1970fr}; later, this method has been adapted to $U(1)$ lattice gauge theory to naturally define Dirac quantization and monopoles \cite{Einhorn:1977jm}, whose significance has been emphasized in the recent years \cite{Sulejmanpasic:2019ytl,Chen:2019mjw}.

In the Villainized $U(1)$ gauge theory, an integer dynamical variable $s_p$ is introduced on the plaquette, so that $F_p=(\mathrm{d}A)_p +2\pi s_p$ where $A_l=-i\ln u_l$ and $\mathrm{d}$ is the lattice exterior derivative is a well-defined \emph{real} variable (the ambiguity in the logarithm can be absorbed by $s_p$). Summing $F_p$ over the plaquettes $p$ on a closed 2d surface, we get $\sum_p F_p/2\pi=\sum_p s_p\in\mathbb{Z}$, the Dirac quantization condition.

\begin{figure}
    \centering
    \includegraphics[width=0.25\linewidth]{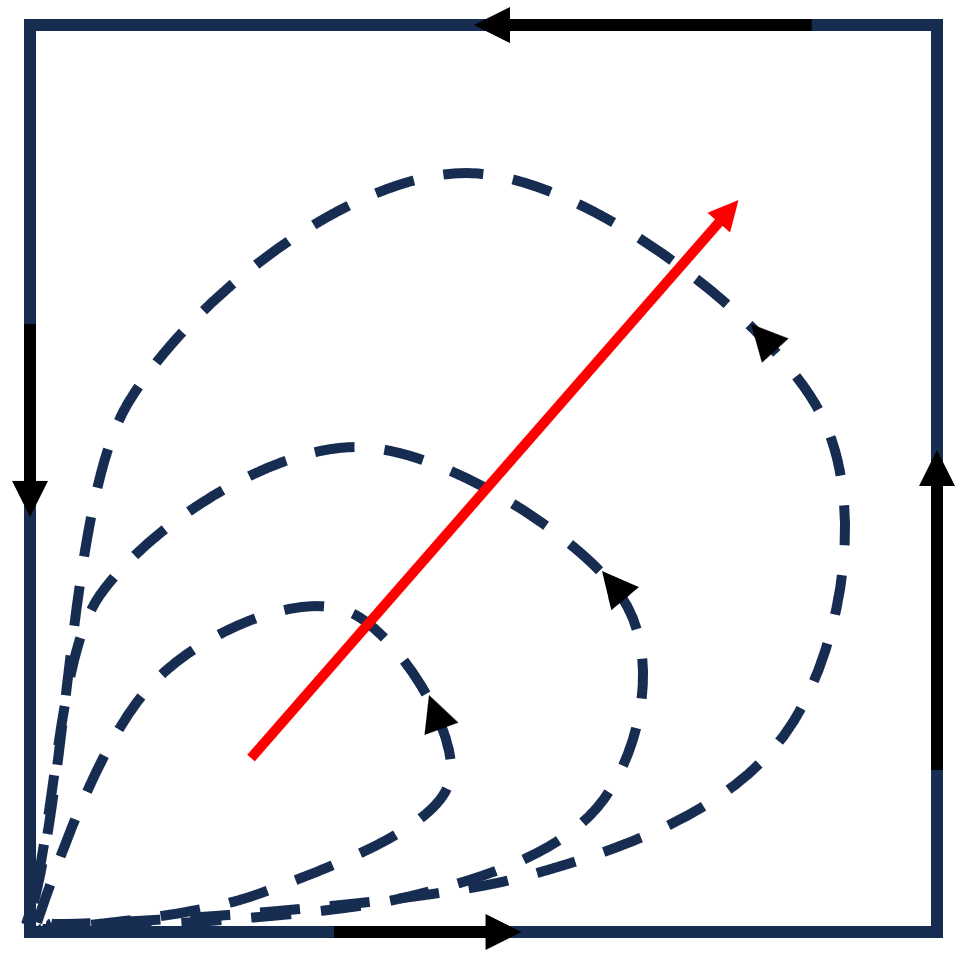}
    \caption{Interpolating the Wilson loop into the plaquette. As the size of the loop grows, the holonomy interpolates from 1 to $u_p$.}
    \label{fig:into_plaq}
\end{figure}

At this point we want to emphasize that, the fundamental reason why the $U(1)$ gauge flux over the plaquette, $F_p$, is real, is \emph{not} because we want it to be a Lie algebra element---this perspective would be too ``perturbative''. Rather, it is because we want to consider how the holonomy around a plaquette might interpolate \emph{into} the plaquette---a topological (homotopy theoretic) perspective. Think of the plaquette as embedded in the continuum, and then, as in Fig.~\ref{fig:into_plaq}, started with a trivial Wilson loop, as the loop grows in size until it matches the plaquette, we can ask how the holonomy gradually interpolates from $1$ to $u_p$. The interpolation process is described by a path in $U(1)$, whose (signed) length is a real number $F_p$ with its $U(1)$ part fixed by $u_p$, i.e. $e^{iF_p}=u_p$, but its integer part admits different possibilities that must be sampled in the path integral with suitable weight---in particular, the weight should decrease as $|F_p|$ increases. This is the key idea behind Villainization. See \cite{Chen:2024ddr} for a detailed review of the Villainization method in relation to the present context.

\

Now let us present the overview of how we will construct a refined version of $SU(2)$ gauge theory on the lattice, so that instanton can be naturally defined; generalization to $SU(N)$ will be discussed later. The Villainization example teaches us that we should introduce suitable dynamical degrees of freedom on higher dimensional lattice cells, in order to sample the different possibilities of how the gauge field sampled on the lattice links might interpolate into the continuum, so that from the interpolation we can extract the desired topological quantities. Now the problem is, what is the guiding principle for how the new degrees of freedom should be introduced? While the principle behind Villainization is as simple as $U(1)=\mathbb{R}/2\pi\mathbb{Z}$, capable of capturing the first Chern class, in \cite{Chen:2024ddr} it is shown that to capture the second Chern class of Yang-Mills gauge field, the lattice degrees of freedom must form suitable higher categories that involves certain parts that \emph{cannot} be described by the familiar notions of groups and fibre bundles. Nonetheless, this seemingly formal structure still admits very intuitive physical interpretation that can be easily made sense of, as we shall see now.

Our refined lattice path integral on a 4d Euclidean lattice looks like
\begin{align}
    Z=&\left[\prod_{l'} \int_{SU(2)}\mathrm{d}u_{l'}\right] \left[\prod_{p'} \sum_{\xi_{p'}=\pm}\int_{S^2}\frac{\mathrm{d}\hat{k}_{p'}}{4\pi}\right]\left[\prod_{c'}\int_{-\pi}^{\pi}\frac{\mathrm{d}\alpha_{c'}}{2\pi}\right]\left[\prod_{h'}\sum_{s_{h'}\in\mathbb{Z}}\right] \nonumber \\[.2cm]
    & \ \prod_p W_2(\Tr \, u_p,\xi_p) \prod_c W_3\left(e^{i\alpha_c}\nu^\ast(u_{l\in \partial c}, \xi_{p\in\partial c}, \hat{k}_{p\in\partial c})+c.c.\right)\prod_h W_4\left(q_h=\frac{(\mathrm{d}\alpha)_h}{2\pi} + s_h\right)
\label{eqn:path_int}
\end{align}
where in the first line, aside from the original Wilson line variable $u_l$ on each link $l$, there are also dynamical variables on each plaquette $p$, cube $c$, and hypercube $h$. In the second line, there are suitable weights on each plaquette, cube and hypercube. Now we explain how this path integral is constructed, how each variable is physically interpreted, and what properties the weights should have.

To motivate the construction, we begin with the expression of the instanton number in the continuum (\ref{eqn:q_def}). We divide the manifold into many 4d polyhedra (``patches'') labeled by $a,b,\dots$, and the gauge field $A$ might be under different gauges in each polyhedron $a$, indicated by a subscript $a$. We can express
\begin{equation}
    \mathcal{Q}=\frac{1}{2(2\pi)^2}\Tr(F\wedge F)=\frac{1}{2(2\pi)^2} \mathrm{d}\Tr\left(A\wedge \mathrm{d}A-\frac{2i}{3}A\wedge A\wedge A\right)_a\equiv \frac{1}{2\pi}\mathrm{d}\mathcal{CS}_a,
\end{equation}
where $\mathcal{CS}_a$ is the continuum Chern-Simons (CS) 3-form in the gauge of the $a$ polyhedron. The instanton density integrated over one polyhedron $a$ can be expressed as
\begin{equation}
    Q_a\equiv \int_a \mathcal{Q}=\frac{1}{2\pi}\oint_{\partial a} \mathcal{CS}_a=\frac{1}{2\pi}\sum_b \int_{ba}\mathcal{CS}_a
    \label{eqn:q_CS_cont}
\end{equation}
where $ba$ mean $b\cap a$, oriented ``outwards'' from $a$.
\footnote{Note that $\int_{ba}\mathcal{CS}_{a}\neq -\int_{ab} \mathcal{CS}_{b}$, because $\mathcal{CS}_a$ and $\mathcal{CS}_b$ are in general under different gauges. This is why $Q=\sum_a Q_a$ can be non-zero on a closed 4d manifold; working out the details more carefully allows one to show $Q\in\mathbb{Z}$, see e.g. \cite{Luscher:1981zq,Chen:2024ddr}.}
It is familiar from CS theory that, if we only know the gauge field (but not specifying the gauge) on the boundary $\partial a$ but not in the interior of $a$, then $\oint_{\partial a} \mathcal{CS}$ is only well-defined as a $U(1)$ phase under large gauge transformation; only when we also specify the field in the interior of $a$ (which disables large gauge transformations) will $Q_a$ be well-defined as a real number \cite{Dijkgraaf:1989pz}, such that it will become an integer if $a$ does not have any boundary. This is much like how, in $U(1)$ gauge theory, if we only know the gauge field on the boundary $\partial a$ of a 2d region $a$ but not in the interior of $a$, then $\oint_{\partial a} A$ is only well-defined as a $U(1)$ phase under large gauge transformation; only when we also specify the field in the interior of $a$ (which disables large gauge transformations) will the flux through the region $a$ be a well-defined real number, such that it will be Dirac quantized if $a$ does not have any boundary.
We already learned that, going from the continuum to the lattice, the latter situation is naturally handled by Villainization. Therefore, if we are somehow able to have a $U(1)$ phase on each lattice 3d cube that conceptually represents the continuum CS integral over that cube, then we can use the familiar Villainization procedure to represent instanton density on the 4d hypercubes, which sums up to an integer instanton number.

Let us substantiate the idea. We let $e^{i\alpha_c}\in U(1)$ on each cube $c$ denote the CS phase over that cube---how  $e^{i\alpha_c}\in U(1)$ is related to the $SU(2)$ gauge field on the links will be discussed below, for now let us assume we already have this $U(1)$ variable that adequately represents the CS phase over cube $c$. Then $e^{i(\mathrm{d}\alpha)_h}\in U(1)$ (here $\mathrm{d}$ is the lattice exterior derivative) is the CS phase over the cubes on the boundary of a hypercube $h$. We introduce a dynamical variable $s_h\in \mathbb{Z}$ in each hypercube $h$ as the Villainization variable, so that
\begin{equation}
    q_h=\frac{(\mathrm{d}\alpha)_h}{2\pi}+s_h
    \label{eqn:q_latt}
\end{equation}
\footnote{We can e.g. fix $\alpha_c \in (-\pi, \pi]$, since the $2\pi$ ambiguity can be absorbed by $s_h$, just like in the usual Villain model.}
is physically interpreted as the instanton density integrated over that hypercube, $\int_h \mathcal{Q} \rightsquigarrow q_h$. Different $s_h$ can be interpreted as different ways of interpolating the CS 3-form from the boundary of the hypercube into the interior of the hypercube. When summed over all hypercubes on a 4d Euclidean torus, we have the total instanton number 
\begin{equation}
    Q=\sum_h q_h = \sum_h s_h \in \mathbb{Z} \ .
    \label{eqn:total_Q_latt}
\end{equation}
On each hypercube, we should have a weight, interpreted as the local ``instanton fugacity", that depends on the instanton density on that hypercube
\begin{equation}
    \prod_h W_4\left(q_h\right) \ .
\end{equation}
$W_4$ should be a positive function that decreases with $|q_h|$, and its detailed form should be optimized numerically. In this paper, it is understood that the detailed form (aside from some stated basic properties) of any weight function should be optimized numerically, just like in traditional lattice gauge theory. 

Now the problem has been reduced to how to get a $U(1)$ variable $e^{i\alpha_c}$ to adequately represent the CS phase over a cube $c$. The immediate question is, should $e^{i\alpha_c}$ be a function of the gauge field $u_l$ on the links around the cube $c$, or should it be an independent dynamical variable? It turns out it should be an independent dynamical variable: Because $\alpha_c$ is supposed to represent the continuum CS integral of the gauge field over the cube $c$, but given the gauge field on the links around a cube, obviously the interpolation into the cube is non-unique, so we want $e^{i\alpha_c}$ to be dynamical to capture such possible fluctuations in the interpolation. We introduce a cube weight to constrain such fluctuations:
\begin{equation}
    \prod_c W_3(e^{i\alpha_c}\nu^\ast_c + c.c.)
    \label{eqn:W3}
\end{equation}
where $W_3$ is a positive and increasing function, and $\nu_c$ is a complex function that depends on the gauge fields on the boundary of the cube. Its role is so that the phase $\nu_c/|\nu_c|\equiv e^{i\alpha^{(0)}_c}$ is the saddle point for $e^{i\alpha_c}$, hence $W_3$ \emph{probabilistically relates} the dynamical CS phase $e^{i\alpha_c}\in U(1)$ to the dynamical $SU(2)$ gauge field sampled on the links. On the other hand, the norm $|\nu_c|$ is the ``susceptibility'' or ``sensitivity'' of the weight to the dynamical CS phase. This idea is inspired by the familiar spinon decomposition of $S^2$ non-linear sigma model on lattice \cite{Rabinovici:1980dn, DiVecchia:1981eh, sachdev1990effective}, see \cite{Chen:2024ddr} for explanation.

The CS phase saddle $e^{i\alpha^{(0)}_c}$ will be constructed by a ``standard choice of interpolation'' of the gauge field from the boundary of a cube into the interior of the cube, inspired by \cite{Luscher:1981zq}, and the fluctuation of $e^{i\alpha_c}$ away from the saddle point $e^{i\alpha^{(0)}_c}$ captures essential information of how an actual interpolation into the cube might have deviated from this most probable ``standard choice of interpolation''. Notably, the ``standard choice'' cannot be uniquely and continuously made over the entire space of the gauge field configurations on the boundary of the cube---a reminiscence of the original problem (\ref{eqn:origin}). When the ``standard choice'' runs into ambiguity (or say singularity), we require the norm
\begin{equation}
    |\nu_c| \rightarrow 0 \ \ \ \ \text{ as standard interpolation becomes ambiguous}
\end{equation}
so that the cube weight becomes insensitive to $e^{i\alpha_c}$ (as it should intuitively be), and thus the path integral weight does not develop any physical ambiguity/singularity. See \cite{Chen:2024ddr} for a discussion of how these features are in parallel to what happens in the familiar spinon decomposition of $S^2$ non-linear sigma model on lattice \cite{Rabinovici:1980dn, DiVecchia:1981eh, sachdev1990effective}.

The construction of the function $\nu_c$ in Section \ref{s_CS} is the main technical part of this paper. But before that, we must first clarify in what variables the function $\nu_c$ is. Naively, one might think $\nu_c$ is a function $\nu_c=\nu(u_{l\in \partial c})$ of those gauge fields $u_l$ on the links $l$ around the cube $c$. However, a closer thought suggests it should be more than that. We said $\nu_c$ will be constructed by a ``standard interpolation'' of the gauge field into the interior of the cube, given the gauge field on the boundary of the cube, but the boundary of a cube not only consists of links, there are also plaquettes, so we might as well need to first consider the possible interpolations of the gauge fields on the links into the plaquettes. And this is indeed the case.
\begin{figure}
    \centering
    \includegraphics[width=0.3\linewidth]{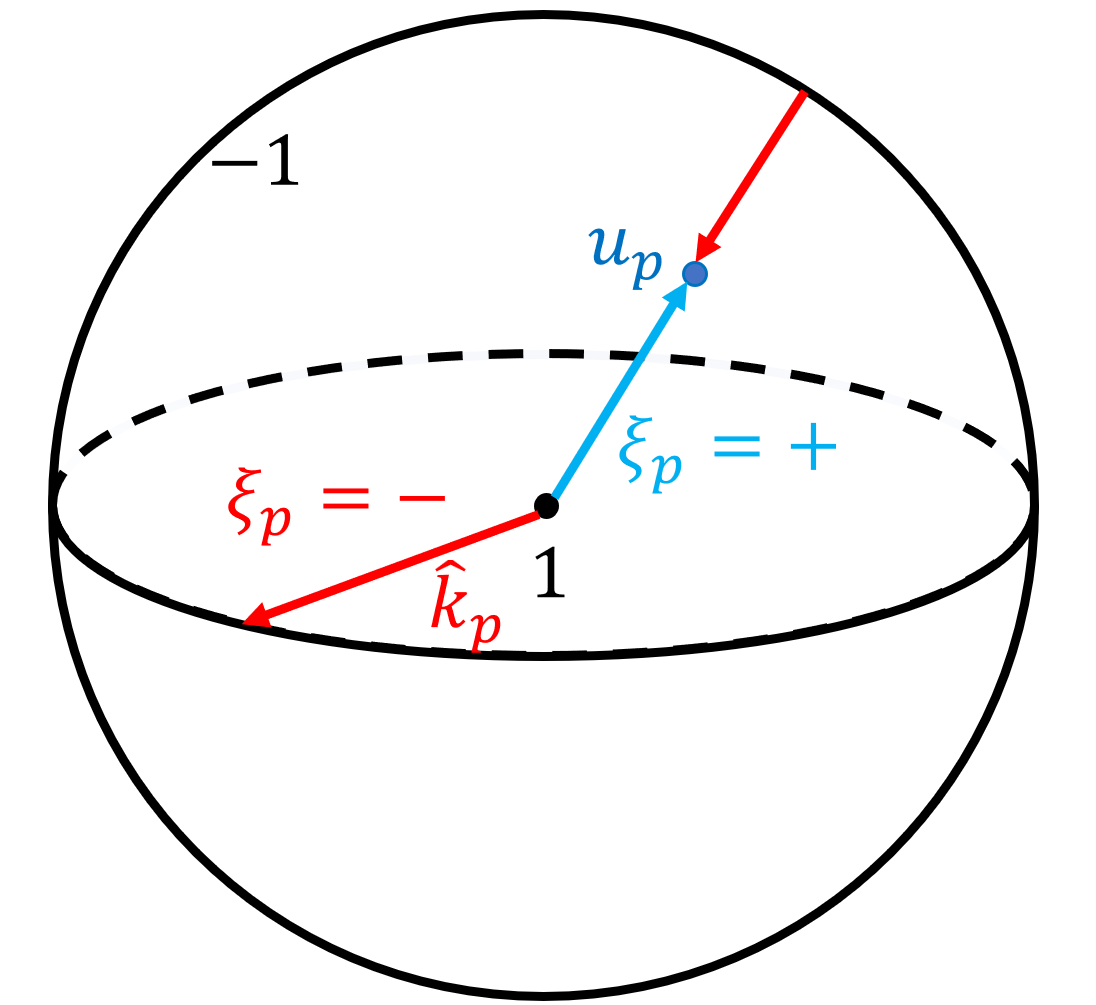}
    \caption{The two ways of interpolation. The solid ball represents $SU(2)$, with the center of the ball representing $1$ and the entire boundary representing $-1$ (hence topologically $S^3$). From this illustration we can clearly see the singularity in each way of interpolation, which is $u_p=-1$ for $\xi_p=+$ and $u_p=1$ for $\xi_p=-$.}
    \label{fig:g_sketch}
\end{figure}

\begin{figure}
    \centering
    \includegraphics[width=0.5\linewidth]{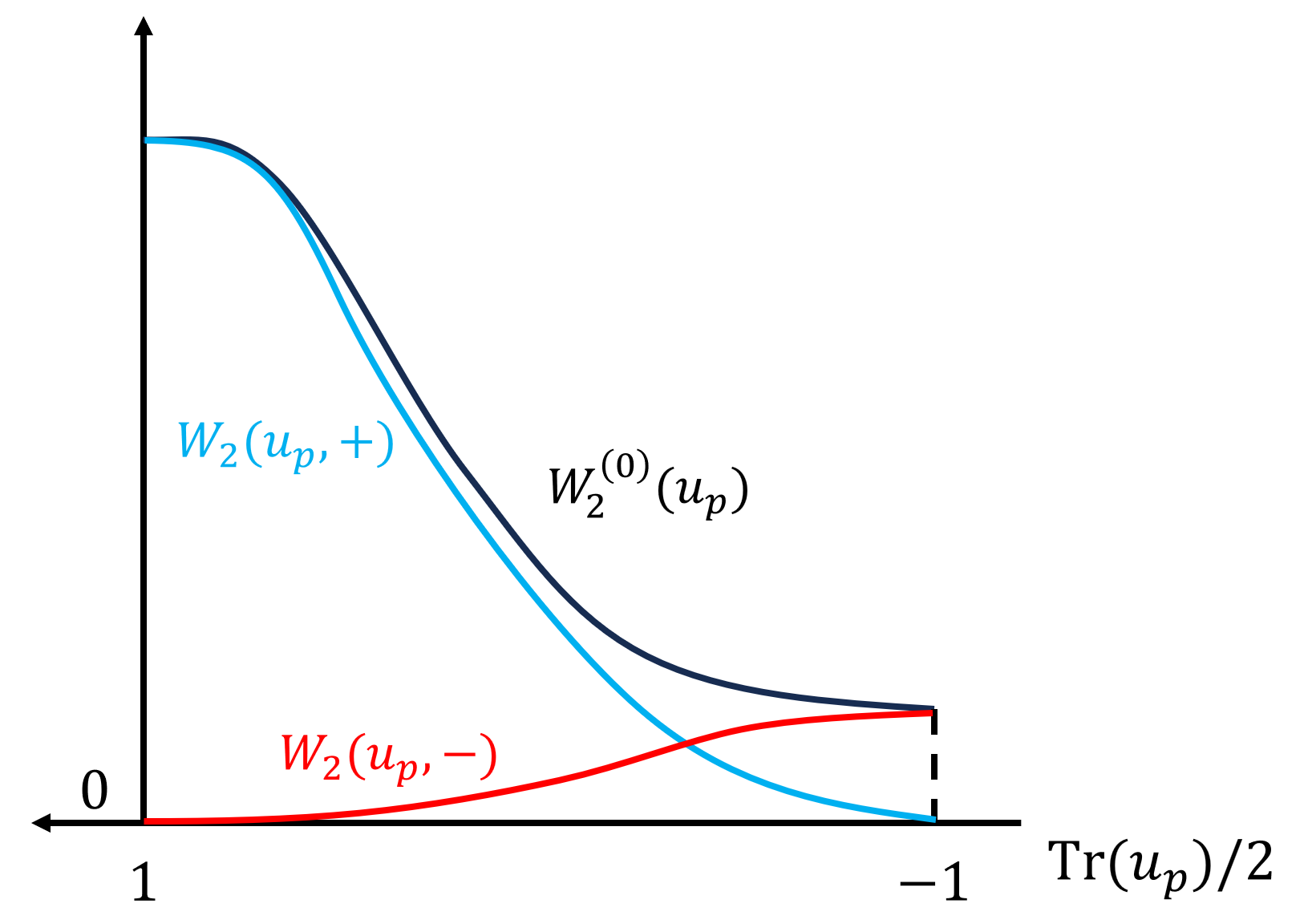}
    \caption{The plaquette weight for two kinds of interpolations, where $W^{(0)}_2$ is the plaquette weight in Wilson's lattice gauge theory.}
    \label{fig:W2_idea}
\end{figure}

The key feature of the categorical construction is that, the plaquette degrees of freedom that samples different ways of interpolating the Yang-Mills gauge field on the links into the plaquette \emph{cannot} be adequately described in terms of groups or fibre bundles, due to certain mathematical constraint \cite{Chen:2024ddr}. But it is not hard to describe them explicitly. Think again about Fig.~\ref{fig:into_plaq}, but now for $SU(2)$ gauge field. It turns out that, we need to sample two kinds of interpolations. The first kind is, as the Wilson loop becomes larger, the holonomy interpolates along the geodesic from $1$ to $u_p$, as illustrated by the blue segment in Fig.~\ref{fig:g_sketch}. The second kind is, the holonomy first interpolates along some direction, labeled as $\hat{k}_p$, from $1$ to $-1$, and then along the geodesic from $-1$ to $u_p$, as illustrated by the red segments in Fig.~\ref{fig:g_sketch}. We can use $\xi_p= \pm$ to label the two kinds of interpolation, and when $\xi_p=-$, we also have $\hat{k}_p\in S^2$ to play the role described above. The $\xi_p$ and $\hat{k}_p$ on each plaquette are dynamical variables in the path integral. The $\nu_c$ in the cube weight (\ref{eqn:W3}) will then be a function of the link and plaquette variables around the cube:
\begin{equation}
    \nu_c=\nu(u_{l\in \partial c}, \xi_{p\in\partial c}, , \hat{k}_{p\in\partial c}) \ .
\end{equation}
Moreover, on each plaquette there should be a plaquette weight for the two kinds of interpolations,
\begin{equation}
    \prod_p W_2(\Tr \, u_p,\xi_p)
\end{equation}
that looks like Fig.~\ref{fig:W2_idea} (note $\hat{k}_p$ only appears in $W_3$ but not in $W_2$). Note that we have written the argument as $\Tr \, u_p$ so that the weight is gauge invariant. A crucial property of $W_2$ is that
\begin{equation}
    W_2(\Tr \, u_p=-2, \xi_p=+)=0, \ \ \ \ \ 
    W_2(\Tr \, u_p=2, \xi_p=-)=0.
\end{equation}
This is obviously because when $\xi_p=+$, we are interpolating from $1$ to $u_p$ along the geodesic, but the direction of the geodesic will become ambiguous when $u_p\rightarrow -1$; likewise when $\xi_p=-$.
\footnote{One might wonder why, in $W_2$ depicted by Fig.~\ref{fig:W2_idea}, when $u_p$ is near $-1$, the second way of interpolation can have higher weight than the first way of interpolation. We can think of each of these two ways of interpolations as a representative for a class of generic interpolations. Then we think of the weight as the exponentiation of the free energy, so we are not only comparing the energies of the representative paths, but also the entropies in the classes they represent \cite{Chen:2024ddr}.}

If we temporarily ignore the weights $W_3, W_4$ and look at $W_2$ only, summing up different plaquette interpolations should recover Wilson's plaquette weight (\ref{eqn:wilson_weight}):
\begin{equation}
    \int_{S^2}\frac{\mathrm{d}\hat{k}_p}{4\pi}\sum_{\xi_p=\pm}W_2(\Tr \, u_p,\xi_p)=\sum_{\xi_p=\pm}W_2(\Tr \, u_p,\xi_p) \simeq W_2^{(0)}(\Tr \, u_p).
    \label{eqn:W2_recovery}
\end{equation}
(Here ``$\simeq$'' means it is unimportant for the functions to be exactly equal---though we can make them so if we really want to---because the detailed forms of the weights are not of fundamental importance, rather they are to be optimized numerically anyways.) In the actual construction $W_3$ and $W_4$ are non-trivial, so summing/integrating over the plaquette variables as well as the cube and hypercube variables will generate effective couplings of $u_p$ between different plaquettes. It is reasonable to expect that, as long as the detailed parameters in the weights are chosen within a suitable range (which can only be determined numerically), the dynamics would not be significantly different from (\ref{eqn:wilson_weight}). In fact, we may even hope for these not-exactly-local effective couplings to be helpful, as they might effectively play some role of Symanzik improvement \cite{Symanzik:1983dc, Weisz:1982zw, Curci:1983an}---a point that we will come back to at the end of the paper. 

As intuitive as this is, the $\xi_p$ degree of freedom does not form a group in any sense (in particular it is not a $\mathbb{Z}_2$ variable), because composition is meaningless; moreover, the $\xi_p$ and $u_p$ together do not form a fibre bundle over $SU(2)$, because $u_p=\pm 1\in SU(2)$ are special points as they are excluded by $\xi_p=\mp$ respectively, while in a fibre bundle the base manifold should have no special point.
\footnote{The $(u_p, \xi_p, \hat{k}_p)$ on a plaquette forms the space $SU(2)\backslash \{-1\}\sqcup \left( SU(2)\backslash\{-1\} \times S^2 \right)$, because $\xi_p=+$ excludes $u_p=-1$, while $\xi_p=-$ excludes $u_p=1$ and at the same time makes $\hat{k}_p$ meaningful. This space is a submersive cover of, but not a fibre bundle over, $SU(2)$.}
Mathematically, these seemingly ``not so nice'' features are unavoidable when trying to implement the second Chern class (and most generic topological classes) on the lattice using finite dimensional local degrees of freedom \cite{Chen:2024ddr}.

Summarizing everything above, we arrive at the refined path integral (\ref{eqn:path_int}):
\begin{align*}
    Z=&\left[\prod_{l'} \int_{SU(2)}\mathrm{d}u_{l'}\right] \left[\prod_{p'} \sum_{\xi_{p'}=\pm}\int_{S^2}\frac{\mathrm{d}\hat{k}_{p'}}{4\pi}\right]\left[\prod_{c'}\int_{-\pi}^{\pi}\frac{\mathrm{d}\alpha_{c'}}{2\pi}\right]\left[\prod_{h'}\sum_{s_{h'}\in\mathbb{Z}}\right] \nonumber \\[.2cm]
    & \ \prod_p W_2(\Tr\, u_p,\xi_p) \prod_c W_3\left(e^{i\alpha_c}\nu^\ast(u_{l\in \partial c}, \xi_{p\in\partial c}, \hat{k}_{p\in\partial c})+c.c.\right) \prod_h W_4\left(q_h=\frac{(\mathrm{d}\alpha)_h}{2\pi} + s_h\right).
\end{align*}
In the first line, we have the original $SU(2)$ Wilson line link variable $u_l$, the necessarily-non-group-valued plaquette variable $\xi_p$ and $\hat k_p$, the CS phase $U(1)$ cube variable $\alpha_c$, and the Villainization hypercube variable $s_h$. (Altogether these degrees of freedom form a higher category, more exactly a suitable weak 4-group that ``implements'' the second Chern class of $SU(2)$ in some precise mathematical sense. See \cite{Chen:2024ddr} for the mathematical framework.) In the second line we have the plaquette weight $W_2$ that looks like Fig.~\ref{fig:W2_idea}, the cube weight $W_3$ that is positive and increasing with the argument and hence maximized at the saddle $e^{i\alpha_c}=e^{i\alpha^{(0)}_c}\equiv \nu_c/|\nu_c|$, and the hypercube weight $W_4$ that is positive and decreasing with the magnitude of the argument. Other than these general properties, the detailed forms of the weights cannot be inferred from fundamental considerations, and should be optimized numerically. In particular, the hypercube weight directly controls the local instanton fugacity, and the total instanton number is $Q=\sum_h q_h=\sum_h s_h$.

In practice, it is much more convenient to use the ``CS phase deviation" $e^{i\beta_c}\equiv e^{i\alpha_c-i\alpha_c^{(0)}}=e^{i\alpha_c} \nu_c^\ast/|\nu_c|$ as the dynamical $U(1)$ cube variable, so that the path integral becomes
\begin{align}
    Z=&\left[\prod_{l'} \int_{SU(2)}\mathrm{d}u_{l'}\right] \left[\prod_{p'} \sum_{\xi_{p'}=\pm}\int_{S^2}\frac{\mathrm{d}\hat{k}_{p'}}{4\pi}\right]\left[\prod_{c'}\int_{-\pi}^{\pi}\frac{\mathrm{d}\beta_{c'}}{2\pi}\right]\left[\prod_{h'}\sum_{s_{h'}\in\mathbb{Z}}\right] \nonumber \\[.2cm]
    & \ \prod_p W_2(\Tr \, u_p,\xi_p) \prod_c W_3\left(2|\nu_c| \cos\beta_c \right) \prod_h W_4\left(q_h=\frac{(\mathrm{d}\beta)_h+(\mathrm{d}\alpha^{(0)})_h}{2\pi} + s_h\right).
    \label{eqn:path_int_convenient}
\end{align}
In this form, we no longer need to explicitly construct the CS phase saddle $\alpha^{(0)}_c \mod 2\pi$ on each single cube, which is $SU(2)$ gauge dependent, but only the total CS phase saddle $(\mathrm{d}\alpha^{(0)})_h \mod 2\pi$ around the boundary of each hypercube, which is $SU(2)$ gauge invariant, making the theory better behaved in practical implementation, as we will explain in details. When we construct $(\mathrm{d}\alpha^{(0)})_h \mod 2\pi$ in Section \ref{s_CS}, we will borrow some ideas and results from L\"{u}scher's geometrical construction \cite{Luscher:1981zq} (but with some crucial rewriting), and it will become obvious that the geometrical construction can be interpreted as taking the saddle point approximation for $\xi_p, \hat{k}_p, e^{i\alpha_c}, s_h$ in the fully-fledged categorical construction, restricted to when $u_p$ is close to $1$ (recall the geometrical construction indeed requires discarding the configuration space when any $u_p$ becomes not so close to $1$ \cite{Luscher:1981zq, Phillips:1986qd}).
\footnote{There has also been a model \cite{Seiberg:1984id} that is much like the geometrical construction \cite{Luscher:1981zq}, but having the dynamical CS $U(1)$ phase on the cube; on the other hand, there is no dynamical plaquette and hypercube variables, and $|\nu|$ is constant rather than a function which may, crucially, vanish in certain cases.}

So far we have only discussed pure Yang-Mills theory. This is because the refinement is only needed for the self dynamics of the Yang-Mills gauge field. If quarks are introduced, the gauge field's coupling to quarks stays the same as in Wilson's lattice gauge theory.

Before we proceed, we remind that there are also some important variants of the path integral (\ref{eqn:path_int}) \cite{Chen:2024ddr}:
\begin{itemize}
\item 
We can include a topological theta term $e^{i\Theta Q}=e^{i\Theta\sum_h q_h}=e^{i\Theta \sum_h s_h}$ (although the complex weight makes it hard for Monte-Carlo), with $\Theta$ manifestly $2\pi$ periodic. 

We can even promote $\Theta$ to a dynamical axion field $\theta_h$, with coupling $e^{i\sum_h\theta_h q_h + \sum_c im_c \alpha_c}$, where $m_c\in \mathbb{Z}$ is the dynamical Villainization part of $\theta_h$ (viewed on the dual lattice) so that the local $2\pi$ ambiguity in $\theta_h$ can be absorbed by $m_c$. The axion field has its own dynamical weight, as a function of $(\mathrm{d}^\star\theta+2\pi m)_c$ on each cube $c$, where $\mathrm{d}^\star$ is the dual lattice exterior derivative.

\item 
In 3d spacetime instead of 4d, we can have (\ref{eqn:path_int}) without the hypercube degree of freedom and hypercube weight, and moreover include an extra complex weight $e^{iK\sum_c\alpha_c}$. This is the natural implementation of the lattice Chern-Simons-Yang-Mills theory. The CS level $K$ is manifestly required to be an integer. Although this cannot be studied via Monte-Carlo due to the complex phase, nor is it solvable like the $U(1)$ case \cite{Peng:2024xbl, Xu:2024hyo}, it is nonetheless conceptually important to have such a mathematically natural lattice implementation. It will be interesting to analyze this theory in the future.

\item
In 5d spacetime or above, the Yang monopole defect \cite{Yang:1977qv} is defined on the lattice as $\mathrm{d}q=\mathrm{d}s\in\mathbb{Z}$ on 5d hypercube. We may include a fugacity weight for the Yang monopole defect, or forbid it by introducing a dynamical $U(1)$ Lagrange multiplier which manifests the associated $(d-5)$-form $U(1)$ global symmetry (on the dual lattice).
\end{itemize}

\section{Construction of the Chern-Simons Saddle around a Hypercube}
\label{s_CS}

In this section we will explain the technical construction of the CS phase saddle $e^{i(\mathrm{d}\alpha^{(0)})_h}$ on the cubes around each hypercube, as well as the CS sensitivity $|\nu_c|$ on each cube (see (\ref{eqn:path_int_convenient})). The essential construction of $(\mathrm{d}\alpha^{(0)})_h$ (up to $2\pi\mathbb{Z}$) follows L\"{u}scher's geometrical construction \cite{Luscher:1981zq}, which involves a standard choice of interpolations of the Wilson loops into the interior of each cube. However, there are some crucial conceptual shifts:
\begin{enumerate}
    \item The Wilson loop $u_p$ around each plaquette is required to be sufficiently close to $1$ in the geometrical construction \cite{Luscher:1981zq, Phillips:1986qd}, hence a large portion of the configuration space is forbidden. While in the full categorical construction, $u_p$ can be any group element.
    \item Before the Wilson loops are interpolated into the interior of each cube, they are first interpolated into the interior of each plaquette. The geometrical construction \cite{Luscher:1981zq} fixes the choice of plaquette interpolation, while in the full categorical construction the plaquette interpolation admits different choices, as explained by Fig.~\ref{fig:g_sketch} and Fig.~\ref{fig:W2_idea}. The choice in \cite{Luscher:1981zq} corresponds to the $\xi_p=+$ choice for $u_p$ close to $1$.
    \item The geometrical construction \cite{Luscher:1981zq} also fixes the choice of cube interpolation, while in the full categorical construction, we can think of the ``actual cube interpolation'' as fluctuating around this standard choice of cube interpolation, with the deviation effectively captured by the CS phase deviation $e^{i\beta_c}$ in (\ref{eqn:path_int_convenient}). The fluctuation is weighted by $W_3$ and the $|\nu_c|$ coefficient, whose required properties will be explained.
    \item In \cite{Luscher:1981zq}, $(\mathrm{d}\alpha^{(0)})_h$ is constructed as a real number. We will show its $2\pi\mathbb{Z}$ part is actually gauge dependent while its $U(1)$ part is gauge invariant. 
    (When $u_p$ is sufficiently close to $1$, as is indeed demanded in \cite{Luscher:1981zq}, the dependence of the $2\pi\mathbb{Z}$ part on gauge choice drops.) We will only use its gauge invariant $U(1)$ part, while the integer part $s_h$ of $q_h$ in (\ref{eqn:path_int_convenient}) is dynamical in the path integral---which can be interpreted as sampling the possible interpolations of the gauge field into the interior of the hypercube.
    \item Moreover, it is important to rewrite the gauge invariant $U(1)$ part $(\mathrm{d}\alpha^{(0)})_h \mod 2\pi$ into a \emph{manifestly} gauge invariant form, so that no gauge choice in the interior of the plaquettes and cubes needs to be made in the intermediate steps of the protocol, in contrast to \cite{Luscher:1981zq}, and this will avoid those ``unnecessary singularities'' in the protocol which are actually due to choosing gauges.
\end{enumerate}
(We can also directly construct the CS phase saddle $e^{i\alpha^{(0)}_c}$ on each single cube, as opposed to $e^{i(\mathrm{d}\alpha^{(0)})_h}$ on the cubes around a hypercube. This is necessary for constructing the 3d Chern-Simons-Yang-Mills theory mentioned at the end of the previous section, as there is no hypercube on 3d lattice. We will leave the direct construction of $e^{i\alpha^{(0)}_c}$ to Appendix \ref{app:CS_cube}, since we do not need it for our main application to 4d lattice QCD.)

\subsection{Interpolation of Wilson Loops into the Cube}
\label{ss_interpolation}

As we have argued in Section \ref{s_path_int}, the main technical issue is to obtain the CS phase saddle $e^{i(\mathrm{d}\alpha^{(0)})_h}$. Following the idea in \cite{Luscher:1981zq}, to achieve this goal, we shall appropriately interpolate the gauge field into the interior of the cube.

First we present our notation on the square lattice on a 4d Euclidean spacetime manifold. Each hypercube $h$ is marked by a vertex $n\in \mathbb{Z}^4$ (up to some periodic boundary condition), such that
\begin{equation}
    h(n)\equiv \{x \in \mathbb{R}^4 \: | \: 0\leq (x^\mu-n^\mu)\leq 1 \quad \forall \mu\in\{1,2,3,4\}\}.
\end{equation}
The hypercubes intersect at cubes
\begin{equation}
    c(n,\mu)\equiv \{x\in h(n)|x^\mu=n^\mu\}=h(n)\cap h(n-\hat{\mu}).
\end{equation}
And the cubes intersect at plaquettes
\begin{equation}
    p(n,\mu,\nu)\equiv \{x\in h(n)|x^\mu=n^\mu,x^\nu=n^\nu\}\quad(\mu\neq \nu).
\end{equation}
If two vertices $n$ and $m$ are connected by a link $\langle nm\rangle$, then we have the gauge field $u_{nm}=u_{mn}^{-1}$ on the link, such that $u_{nm}$ is the Wilson line from $m$ to $n$. Generally, we use $l$ to denote oriented links.

Now we explain what is meant by ``appropriately'' at the beginning of this subsection. This means we want our interpolation protocol to be gauge independent, and therefore the protocol should be describable in terms of Wilson loops only (as opposed to Wilson lines with distinct end points); moreover, we shall choose the starting points of such Wilson loops at the lattice vertices, and ensure the protocol to have the property that under $SU(2)$ gauge transformation at the vertices, the intended interpolated Wilson loops \emph{indeed} transform by conjugation accordingly.
\footnote{In \cite{Luscher:1981zq}, Wilson lines with one end at the lattice vertex and another end in the interior of the cube are being constructed; certain gauge choices in the interior of the cube are essentially made during the procedure. In Appendix \ref{app:q_density}, we explain how to turn the protocol described in \cite{Luscher:1981zq} into the manifestly gauge invariant protocol we describe in the main text. This will also show why the constructed $(\mathrm{d}\alpha^{(0)})_h$ has a gauge invariant $U(1)$ part and a gauge dependent $2\pi\mathbb{Z}$ part.}

In the following we will construct the interpolation of certain Wilson loops, including the loops extended into plaquettes and the ones into cubes.

\subsubsection*{Interpolation into the interior of a plaquette}
\label{sec:interpolation_plaquette}
\begin{figure}
    \centering
    \includegraphics[width=0.35\linewidth]{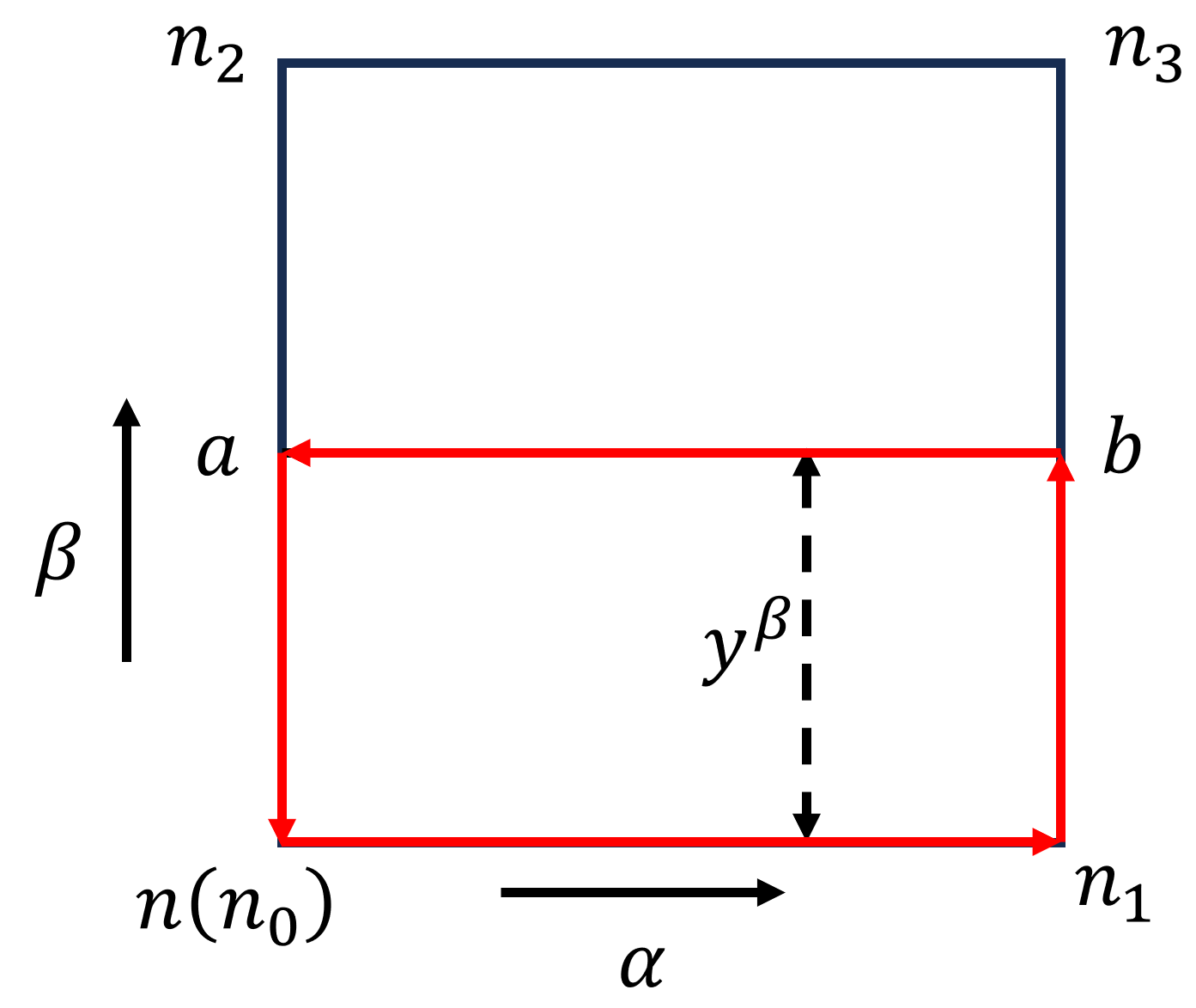}
    \caption{Interpolation on the plaquette $p(n,\mu,\nu) \ (n=n_0)$. We arrange the axes $\alpha$ and $\beta$ so that $\alpha<\beta\in \{1,2,3,4\}\backslash\{\mu,\nu\}$. We want to interpolate the red Wilson loop $(0ab10)$.}
    \label{fig:interpolate_plaq}
\end{figure}

The idea of interpolating the Wilson loop around a plaquette into the interior of the plaquette has already been discussed in Section \ref{s_path_int}, pictured by Fig.~\ref{fig:into_plaq} and Fig.~\ref{fig:g_sketch}. Now we introduce the more detailed protocol. Consider Fig.~\ref{fig:interpolate_plaq}, we want to define the red Wilson loop $(0ab10)$. (In the below we will often denote a Wilson line made of straight lines $a\rightarrow b \rightarrow c \rightarrow \dots \rightarrow d\rightarrow e$ as $(ed\dots cba)$.) We introduce a $g$ function which will be constructed below, and let
\begin{equation}
    (0ab10)\equiv g((02310),y^\beta).
\end{equation}
By its physical meaning we require that
\begin{equation}
    g(u,0)=1,\quad g(u,1)=u.
    \label{eqn:g_condition_1}
\end{equation}
Also, if we do a gauge transformation $w$ on vertices, then the interpolated holonomy should be compatible with the gauge transformation on $p=p(n,\mu,\nu)$:
\begin{equation}
    g(u'_p,y^\beta)=w(n)g(u_p,y^\beta)w(n)^{-1},\quad u'_p=w(n)u_p w(n)^{-1},
    \label{eqn:g_condition_2}
\end{equation}
where $u_p$ is the honolomy around plaquette $p$.

As we have sketched in Fig.~\ref{fig:g_sketch}, we actually need two choices, $g_{\xi_p}$ for $\xi_p=\pm$. Let $u_p=e^{i\vec{t}_p\cdot\vec\sigma}$ with $0\leq|\vec{t}_p|\leq \pi$. Then the simple $\xi_p=+$ choice is 
\begin{equation}
    g_+(u_p,y^\beta)=u_p^{y^\beta} \equiv e^{iy^\beta\vec{t}_p\cdot\vec\sigma} ,
\end{equation}
which is not well-defined only for $u_p=-1$ since the direction $\hat{t}_p$ is not well-defined. On the other hand, we choose
\begin{equation}
    g_-(u_p=e^{i\vec t_p\cdot \vec \sigma},y^\beta;\hat{k}_p)=\left\{
    \begin{aligned}
        &e^{i2\pi y^\beta\hat{k}_p\cdot \vec{\sigma}} \quad y^\beta\in\left[0,\frac{1}{2}\right],\\
        &e^{i[\pi-(2y^\beta-1)(\pi-t_p)]\hat{t}_p\cdot \vec{\sigma}} \quad y^\beta\in \left[\frac{1}{2},1\right],
    \end{aligned}
    \right.
    \label{eqn:g-_def}
\end{equation}
where $t_p\equiv |\vec t_p|$ and $|\hat{k}_p|=1$. Note $g_-$ is well-defined for $u_p=-1$ but not for $u_p=1$. In order for $g_-$ to satisfy condition (\ref{eqn:g_condition_2}), under a gauge transformation $\hat{k}_p$ must transform as
\begin{equation}
    \hat{k}_p\cdot \vec\sigma \rightarrow \hat{k}'_p\cdot \vec\sigma=w(n)(\hat{k}_p\cdot \vec\sigma)w(n)^{-1} \quad \mbox{on} \quad p=p(n,\mu,\nu).
    \label{eqn:k_gauge_transf}
\end{equation}
By the definition above, under a gauge transformation $w(n)$, on $p=p(n,\mu,\nu)$ we have:
\footnote{The $g_+$ function is independent of $\hat{k}_p$, but we leave it as an argument to keep the formula compact.}
\begin{equation}
    g_{\xi_p}(u'_p,y^\beta;\hat{k}'_p)=w(n)g_{\xi_p}(u_p,y^\beta;\hat{k}_p)w(n)^{-1},\quad \xi_p=\pm,
    \label{eqn:g_gauge_trf}
\end{equation}
which is exactly what we want in (\ref{eqn:g_condition_2}).

\subsubsection*{Interpolation into the interior of a cube}
\label{sec:interpolation_cube}

\begin{figure}
    \centering
    \includegraphics[width=0.5\linewidth]{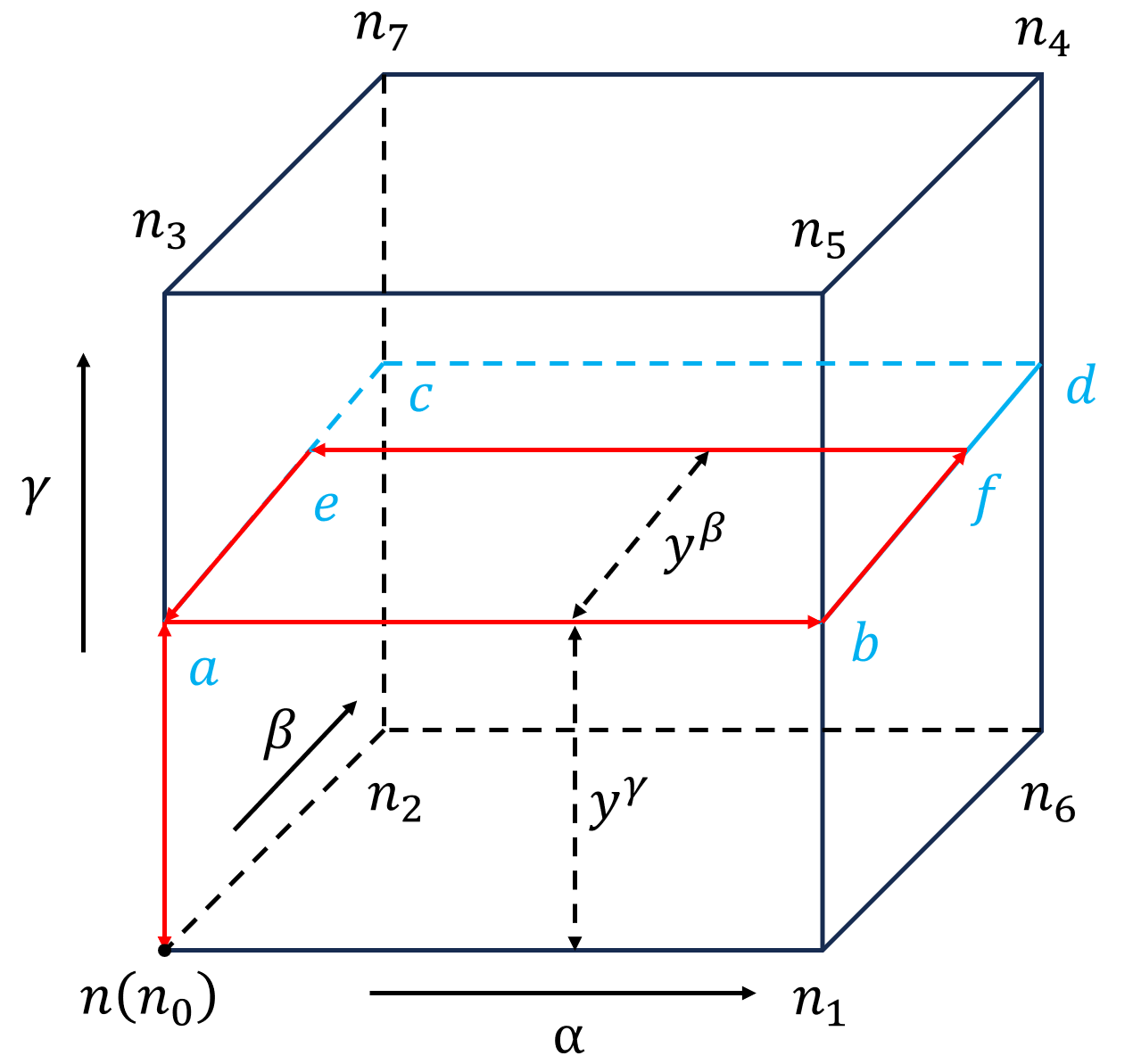}
    \caption{Interpolation on the cube $c(n,\mu) \ (n=n_0)$. We arrange the axes $\alpha$, $\beta$ and $\gamma$ so that $\alpha<\beta<\gamma\in \{1,2,3,4\}\backslash\{\mu\}$. We want to interpolate the red Wilson loop $(0aefba0)$.}
    \label{fig:interpolate_cube}
\end{figure}

On each cube, see Fig.~\ref{fig:interpolate_cube}, we want to construct the red Wilson loop by introducing an $h$ function
\begin{equation}
    (0aefba0)\equiv h((0acdba0),y^\beta,y^\gamma).
\end{equation}
Let us denote $(0acdba0)$ by $u_c(y^\gamma)$, and it is already specified given the $g_\xi$ functions, since
\begin{equation}
    u_c=(0ac20)(02)(2cd62)(26)(16)^{-1}(1bd61)^{-1}(01)^{-1}(0ab10)^{-1}.
    \label{eqn:u_c}
\end{equation}
The physical meaning of the $h$ function requires that
\begin{equation}
    h(u_c(y^\gamma),0,y^\gamma)=1,\quad h(u_c(y^\gamma),1,y^\gamma)=u_c(y^\gamma),
    \label{eqn:h_condition_0}
\end{equation}
\begin{equation}
    h(u_c(0)=(02610),y^\beta,0)=g_{\xi_1}((02610),y^\beta;\hat{k}_1),
    \label{eqn:h_condition_1}
\end{equation}
\begin{equation}
    h(u_c(1)=(0374530),y^\beta,1)=(03)g_{\xi_2}((37453),y^\beta;\hat{k}_2)(30).
    \label{eqn:h_condition_2}
\end{equation}
Here the $\xi_i=\pm$ and $\hat{k}_i \ (i=1,2)$ are those on the bottom plaquette $(02610)$ and the top plaquette $(37453)$ respectively. These requirements suggest $h$ should depend on $\xi_i$ and $\hat{k}_i(i=1,2)$, thus we denote it by $h_{\xi_1\xi_2}(u_c(y^\gamma),y^\beta,y^\gamma;\hat{k}_1,\hat{k}_2)$.

Now consider a gauge transformation $w$ on the vertices. $u_c$ indeed gauge transforms as desired
\begin{equation}
    u'_c(y^\gamma)=w(n_0)u_c(y^\gamma)w(n_0)^{-1}
    \label{eqn:u_c_gauge_trf}
\end{equation}
because $g_\xi$ satisfies the condition (\ref{eqn:g_gauge_trf}). Then, for the $h$ function, we require
\begin{equation}
    h_{\xi_1\xi_2}(u'_c(y^\gamma),y^\beta,y^\gamma;\hat{k}'_1,\hat{k}'_2)=w(n_0)h_{\xi_1\xi_2}(u_c(y^\gamma),y^\beta,y^\gamma;\hat{k}_1,\hat{k}_2)w(n_0)^{-1}
    \label{eqn:h_condition_3}
\end{equation}
where $\hat{k}'_1\cdot\vec\sigma=w(n_0)(\hat{k}_1\cdot\vec\sigma)w(n_0)^{-1}$, $\hat{k}'_2\cdot\vec\sigma=w(n_3)(\hat{k}_2\cdot\vec\sigma)w(n_3)^{-1}$.

We have listed four conditions (\ref{eqn:h_condition_0}), (\ref{eqn:h_condition_1}), (\ref{eqn:h_condition_2}) and (\ref{eqn:h_condition_3}). Now we construct the four $h_{\xi_1\xi_2}$'s.

\vspace{.5cm}\noindent
$\boldsymbol{\xi_1=\xi_2=+.}$ In this case, the function $h$ can simply be chosen as
\begin{equation}
    h_{++}(u_c(y^\gamma),y^\beta,y^\gamma;\hat{k}_1,\hat{k}_2)=g_+\left(u_c(y^\gamma),y^\beta\right)=(u_c(y^\gamma))^{y^\beta}.
\end{equation}
One can easily verify that it satisfies all four conditions. But there is one issue. If there exists a $y^\gamma$ such that $u_c(y^\gamma)=-1$, then $h_{++}$ will be ill-defined for that $y^\gamma$. As discussed in Section \ref{s_path_int}, in this case we should send $|\nu_c|$ to zero. We will discuss more about $|\nu_c|$ in Section \ref{ss_CS_sensitivity}.

\begin{figure}
    \centering
    \includegraphics[width=0.3\linewidth]{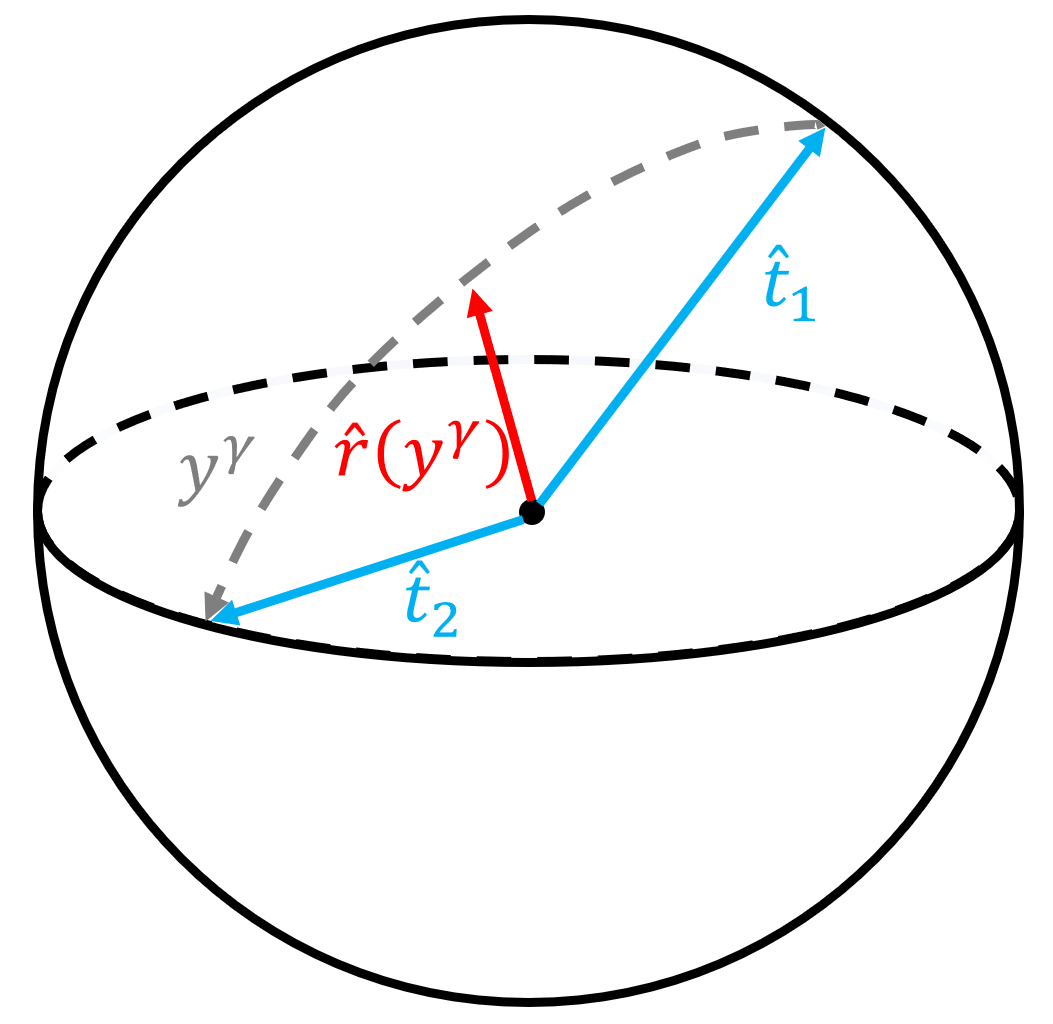}
    \caption{Sketch of the $\hat{r}(y^\gamma;\hat t_1\cdot \vec\sigma,\hat t_2\cdot \vec\sigma)$ function. As $y^\gamma$ varies from 0 to 1, $\hat r(y^\gamma)$ proceeds from $\hat t_1$ to $\hat t_2$ along the geodesic on $S^2$.}
    \label{fig:r_sketch}
\end{figure}

\vspace{.5cm}\noindent
$\boldsymbol{\xi_1=\xi_2=-.}$ Define
\begin{equation}
    h_{--}(u_c(y^\gamma),y^\beta,y^\gamma;\hat{k}_1,\hat{k}_2)=g_-\left(u_c(y^\gamma),y^\beta;\hat r\left(y^\gamma;\hat{k}_1\cdot \vec\sigma,(03)(\hat{k}_2\cdot \vec\sigma)(30)\right)\right)
    \label{eqn:h_--}
\end{equation}
where $\hat r\left(y^\gamma;\hat{k}_1\cdot \vec\sigma,(03)(\hat{k}_2\cdot \vec\sigma)(30)\right)$ is defined as the following. For general $\hat t_1$ and $\hat t_2$, consider
\begin{equation}
    e^{i\vec r(y^\gamma)\cdot\vec\sigma}=\left(e^{i\hat t_2\cdot\vec\sigma}e^{-i\hat t_1\cdot \vec\sigma}\right)^{y^\gamma} e^{i\hat t_1\cdot \vec\sigma}
\end{equation}
and define
\begin{equation}
    \hat{r}(y^\gamma;\hat t_1\cdot \vec\sigma,\hat t_2\cdot \vec\sigma)\equiv \frac{\vec r(y^\gamma)}{|\vec r(y^\gamma)|}.
\end{equation}
Schematically, $\hat{r}(y^\gamma;\hat t_1\cdot \vec\sigma,\hat t_2\cdot \vec\sigma)$ is the interpolation between $\hat t_1$ and $\hat t_2$ along the geodesic on the unit sphere, see Fig.~\ref{fig:r_sketch}. Note $\hat{r}(y^\gamma;\hat t_1\cdot \vec\sigma,\hat t_2\cdot \vec\sigma)$ satisfies
\begin{equation}
\begin{aligned}
    &\hat{r}(0;\hat t_1\cdot \vec\sigma,\hat t_2\cdot \vec\sigma)=\hat t_1,\quad \hat{r}(1;\hat t_1\cdot \vec\sigma,\hat t_2\cdot \vec\sigma)=\hat{t}_2,\\
    &w\left(\hat{r}\left(y^\gamma;\hat t_1\cdot \vec\sigma,\hat t_2\cdot \vec\sigma\right)\cdot \vec\sigma\right)w^{-1}=\hat{r}\left(y^\gamma;w(\hat t_1\cdot \vec\sigma)w^{-1},w(\hat t_2\cdot \vec\sigma)w^{-1}\right)\cdot \vec\sigma.
\end{aligned}
\label{eqn:r_condition}
\end{equation}
because the geodesic between two unit vectors will rotate along with the rotation of the unit vectors (in our case, the ``rotation'' is the gauge transformation Eq.~(\ref{eqn:k_gauge_transf})). By defining $\hat r$ in such a way, one can show that the $h$ function defined in Eq.~(\ref{eqn:h_--}) indeed satisfies all four conditions.

We remark that here we may encounter two types of singularities. The first one is when there exists some $y^\gamma$ such that $u_c(y^\gamma)=1$, which is similar to the $\xi_1=\xi_2=+$ case. The second one is when $\hat{k}_1\cdot \vec\sigma=-(03)(\hat{k}_2\cdot\vec\sigma)(30)$, so that $\hat r\left(y^\gamma;\hat{k}_1\cdot \vec\sigma,(03)(\hat{k}_2\cdot \vec\sigma)(30)\right)$ is ill-defined.
\footnote{This is similar to the spinor-decomposition case when two unit vectors point at opposite directions so that the geodesic between them acquires ambiguity \cite{Chen:2024ddr}}
In both cases we will send $|\nu_c|=0$.

\begin{figure}
    \centering
    \includegraphics[width=1\linewidth]{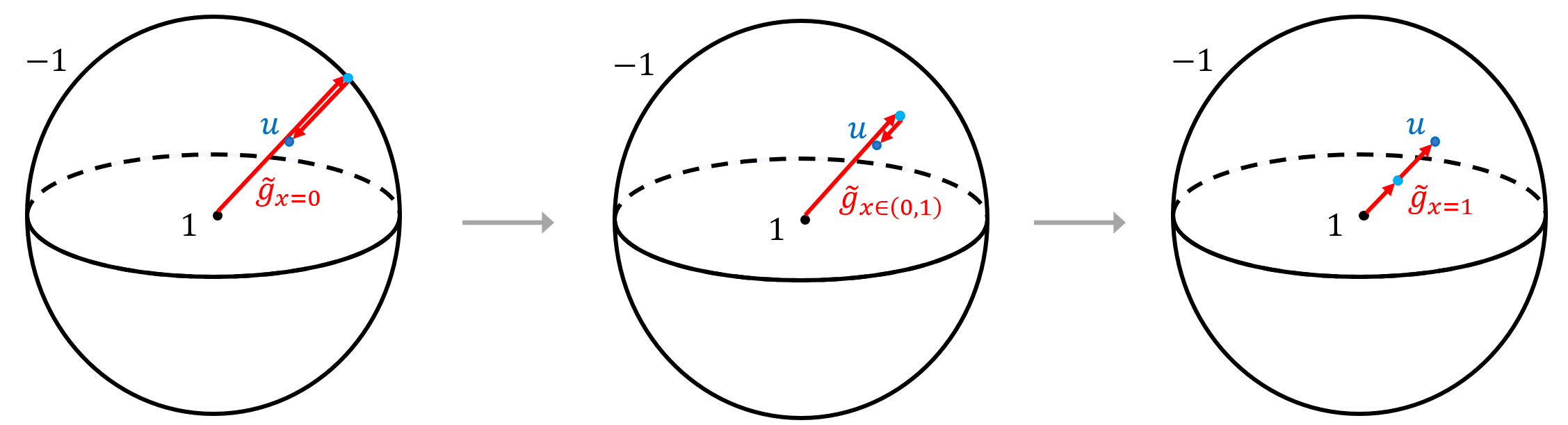}
    \caption{Sketch of the $\tilde g_x$ function. As $x$ increases from 0 to 1, $\tilde g_x$ interpolates from $g_-$ (with $\hat k$ coincides with $\hat t$ of $u$) to $g_+$.}
    \label{fig:g_tilde}
\end{figure}

\vspace{.5cm}\noindent
$\boldsymbol{\xi_1=-,\xi_2=+.}$ Consider the function
\begin{equation}
    h_{-+}(u_c(y^\gamma),y^\beta,y^\gamma;\hat{k}_1,\hat{k}_2)=\left\{
    \begin{aligned}
        &g_-\left(u_c(y^\gamma),y^\beta;\hat{r}(2y^\gamma;\hat{k}_1\cdot \vec\sigma,\hat t_c^{(1/2)}\cdot \vec\sigma)\right) \quad y^\gamma \in \left[0,\frac{1}{2}\right]\\
        &\tilde{g}_{2y^\gamma-1}(u_c(y^\gamma),y^\beta) \quad y^\gamma \in \left[\frac{1}{2},1\right],
    \end{aligned}
    \right.
    \label{eqn:h_construct_-+}
\end{equation}
where $e^{i\vec{t}_c^{(1/2)}\cdot \vec{\sigma}}\equiv u_c(1/2)$ and $\hat t_c^{(1/2)}\equiv \vec t_c^{(1/2)}/|\vec t_c^{(1/2)}|$
so that under a gauge transformation $\hat t_c^{(1/2)'}\cdot \vec\sigma=w(n_0)(\hat t_c^{(1/2)}\cdot \vec\sigma)w(n_0)^{-1}$ by Eq.(\ref{eqn:u_c_gauge_trf}). Recall that $\hat{r}(y^\gamma;\hat{k}_1\cdot \vec\sigma,\hat t_c^{(1/2)}\cdot \vec{\sigma})$ by definition satisfies
\begin{equation}
\begin{aligned}
    &\hat{r}(0;\hat{k}_1\cdot \vec\sigma,\hat t_c^{(1/2)}\cdot \vec\sigma)=\hat{k}_1, \quad \hat{r}(1;\hat{k}_1\cdot \vec\sigma,\hat t_c^{(1/2)}\cdot \vec{\sigma}) =\hat t_c^{(1/2)},\\
    &w(n_0)(\hat{r}(y^\gamma;\hat{k}_1\cdot \vec\sigma,\hat t_c^{(1/2)}\cdot \vec{\sigma})\cdot \vec\sigma)w(n_0)^{-1}=\hat{r}(y^\gamma;\hat{k}'_1\cdot \vec\sigma,\hat t_c^{(1/2)'}\cdot \vec{\sigma})\cdot \vec\sigma.
\end{aligned}
\end{equation}
And $\tilde{g}_x \ (x\in [0,1])$ is defined by
\begin{equation}
    \tilde{g}_x(u=e^{i\vec{t}\cdot \vec{\sigma}},y^\beta)=\left\{
    \begin{aligned}
        &e^{i\left(\pi-\pi x+\frac{t}{2}x\right)2y^\beta \hat{t}\cdot\vec\sigma}\quad y^\beta\in\left[0,\frac{1}{2}\right],\\
        &e^{i\left[t+\left(\pi-\pi x+\frac{t}{2}x - t\right)(2-2y^\beta)\right]\hat{t}\cdot\vec\sigma}\quad y^\beta\in\left[\frac{1}{2},1\right],
    \end{aligned}
    \right.
\end{equation}
as illustrated in Fig.~\ref{fig:g_tilde}, so that
\begin{equation}
\begin{aligned}
    \tilde g_{x=0}(u=e^{i\vec t\cdot \vec \sigma},y^\beta)&=g_-(u,y^\beta;\hat t),\\
    \tilde g_{x=1}(u,y^\beta)&=g_+(u,y^\beta).
\end{aligned}
\end{equation}

By the properties stated above, it can be directly checked that the $h_{-+}$ constructed by (\ref{eqn:h_construct_-+}) satisfies all four conditions.
Actually, although the construction may seem a bit complicated, the idea is simple. 
Going from the bottom to the top of the cube, first we rotate $\hat{k}$ of the function $g_-$ from $\hat k_1$ at the bottom to the direction $\hat t_c^{(1/2)}$ of $u_c(1/2)$, so the $g_-$ function at $y^\gamma=1/2$ is simply an ``extended $u^y$'' (see the left figure of Fig.~\ref{fig:g_tilde}). Then all we have to do is to tune a parameter such that the ``extended $u^y$'' is deformed back to the true $u^y$, which is the $g_+$ on the top plaquette.

As in previous cases, here we may encounter several singularities. The first type of singularity is that $u_c(y^\gamma)=1$ if $y^\gamma\in [0,1/2]$, and similarly $u_c(y^\gamma)=\pm 1$ if $y^\gamma\in [1/2,1]$. The second type is that $\hat t_c^{(1/2)}=-\hat k_1$. We deal with these singularities by setting $|\nu_c|=0$.

\vspace{.3cm}\noindent
$\boldsymbol{\xi_1=+,\xi_2=-.}$ This is similar to the previous case, and we define
\begin{equation}
    h_{+-}(u_c(y^\gamma),y^\beta,y^\gamma;\hat{k}_1,\hat{k}_2)=\left\{
    \begin{aligned}
        &\tilde{g}_{1-2y^\gamma}(u_c(y^\gamma),y^\beta),\quad y^\gamma\in \left[0,\frac{1}{2}\right],\\
        &g_-\left(u_c(y^\gamma),y^\beta;\hat{r}_{+-}\left(y^\gamma;\hat t_c^{(1/2)},\hat{k}_2\right)\cdot \vec{\sigma}\right), \quad y^\gamma\in \left[\frac{1}{2},1\right]
    \end{aligned}
    \right.
    \label{eqn:h_construct_+-}
\end{equation}
where
\begin{equation}
    \hat{r}_{+-}\left(y^\gamma;\hat t_c^{(1/2)},\hat{k}_2\right)\equiv \hat{r}\left(2y^\gamma-1;\hat t_c^{(1/2)}\cdot \vec\sigma,(03)(\hat{k}_2\cdot \vec\sigma)(30)\right).
\end{equation}

\

In summary, we have constructed the interpolations on plaquettes and cubes by introducing the functions $g_\xi(u,y^\beta;\hat{k})$ and $h_{\xi_1\xi_2}(u_c(y^\gamma),y^\beta,y^\gamma;\hat{k}_1,\hat{k}_2)$ respectively. Using these interpolations we will be able to define the CS phase saddle around a hypercube.

\subsection{Expression of the CS Phase Saddle $e^{i(\mathrm{d}\alpha^{(0)})_h}$}
\label{ss_q_formula}

In this subsection, we present the explicit expression of the CS phase saddle around a hypercube using the interpolations we have done in the previous subsection. Our expression is based on a formula derived in \cite{Luscher:1981zq}, see Appendix \ref{app:q_density}, but transformed into a form that is manifestly gauge invariant. More particularly, the original expression in \cite{Luscher:1981zq} involved Wilson lines with end points in the interior of cubes, hence involving gauge choices in the interior (at least in the intermediate steps of the calculation) which can develop artificial gauge choice singularities that cause unnecessary troubles, while we want an expression in terms of the interpolated Wilson loops only, as we argued at the beginning of the last subsection.

\begin{figure}
    \centering
    \includegraphics[width=0.4\linewidth]{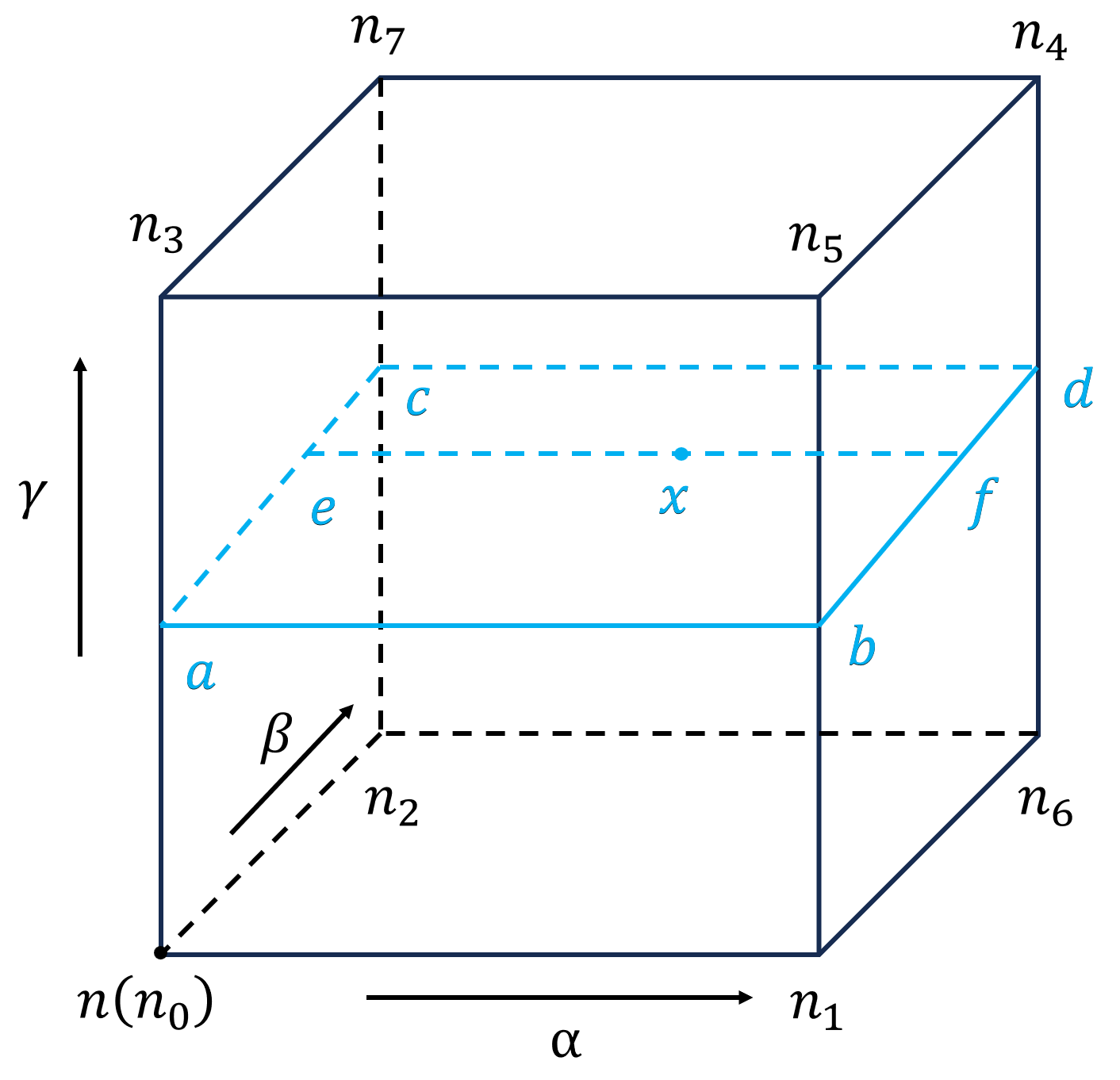}
    \caption{Label of the vertices in a cube $c(n,\mu) \ (n=n_0)$. The coordinate directions $\alpha<\beta<\gamma\in \{1,2,3,4\}\backslash \{\mu\}$.}
    \label{fig:R_def}
\end{figure}

To present our result, given a cube $c(n,\mu)$, we first define a function $R$. In Fig.~\ref{fig:R_def}, for $x\in c(n,\mu) \ (n=n_0)$ we define
\begin{equation}
\begin{aligned}
    R_{n,\mu;\gamma}(x)&\equiv (03),\\
    R_{n,\mu;\beta}(x)&\equiv (0ac2),\\
    R_{n,\mu;\alpha}(x)&\equiv (0aefb1).
\end{aligned}
\label{eqn:R_def}
\end{equation}
In particular, $R_{n,\mu;\nu}(x) \ (\mu\neq \nu)$ is a Wilson line from $n+\hat \nu$ to $n$, running in $c(n,\mu)$, and only depends on those $x^\lambda$ components for $\lambda>\nu$. (In fact, in the below we will only use the $R$ function when $x$ is on the boundary $\partial c(n,\mu)$.) Using the $g$ and $h$ interpolation functions defined in the last subsection, we have the explicit expressions for $R_{n,\mu;\beta}(x)$ and $R_{n,\mu;\alpha}(x)$ (denoting $y\equiv x-n$):
\begin{equation}
\begin{aligned}
    R_{n,\mu;\beta}(x)&= (0ac20)(02)=g_{\xi_p}(u_p,y^\gamma;\hat k_p)(02),\quad p=p(n,\mu,\alpha);\\
    R_{n,\mu;\alpha}(x)&= (0aefba0)(0ab10)(01)\\
    &=h_{\xi_1 \xi_2}(u_c(y^\gamma),y^\beta,y^\gamma;\hat k_1,\hat k_2)g_{\xi_{p'}}(u_{p'},y^\gamma;\hat k_{p'})(01),\quad p'=p(n,\mu,\beta)
\end{aligned}
\end{equation}
where $\xi_i=\pm$ and $\hat{k}_i \ (i=1,2)$ as before label the interpolation chosen on the bottom and top plaquettes respectively, and $u_c(y^\gamma)$ are expressed using $g$ functions according to Eq.(\ref{eqn:u_c}).

\begin{figure}[h]
    \centering
    \includegraphics[width=0.5\linewidth]{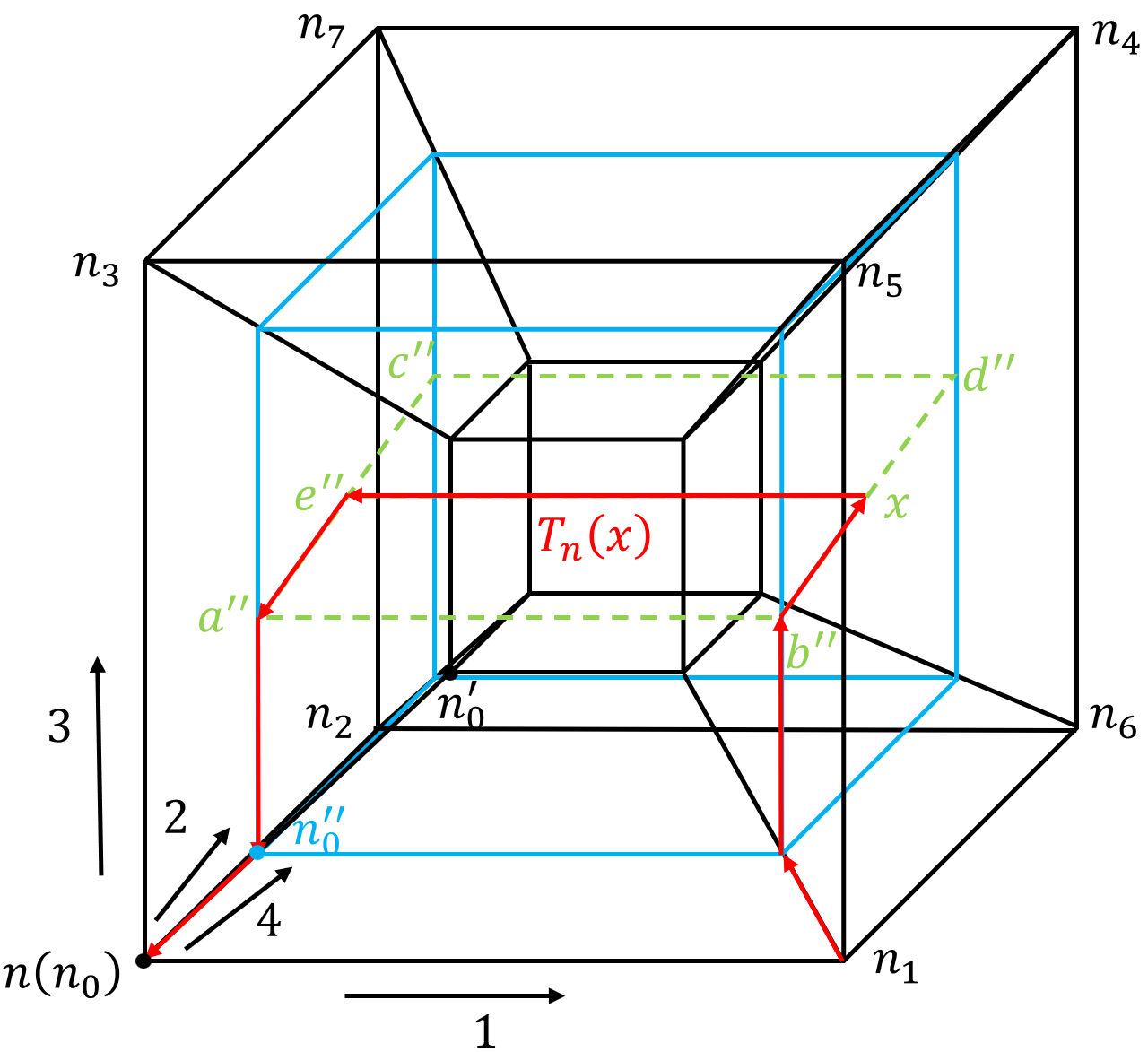}
    \caption{Definition of $T_n(x)$. Here we are (unfortunately) picturing a 4d hypercube $h(n)$, with the outer black cube (on which the vertices are labeled as $n_i$) consisting of points $z$ with $z^4=n^4$, the inner black cube (on which the vertices are labeled as $n'_i$) consisting of points $z$ with $z^4=n^4+1$. The point $x$ is in the cube $c(n+\hat{1},1)$, whose corners are $n_{1,6,5,4}$ and $n'_{1,6,5,4}$; this means $x^1=n^1+1$. The intermediate blue cube (on which the vertices are labeled as $n''_i$) consists of points $z$ with $z^4=x^4$. The red Wilson line from $n_1$ to $n_0$ is the function $T_n(x)$ defined for $x\in c(n+\hat 1,1)$. The Wilson line starts at $z=n_1$, consecutively increases its $z^4, z^3, z^2$ coordinates until matching $x^4, x^3, x^2$ respectively so that $z$ arrives at $z=x$, and then decreases $z^1$ by $1$, and then consecutively decreases its $z^2, z^3, z^4$ until reaching $z=n_0$. }
    \label{fig:T}
\end{figure}

Next, for the hypercube $h(n)$, we introduce the $T$ function. We picture the 4d hypercube $h(n)$ as Fig.~\ref{fig:T}, and consider a point $x$ in the cube $c(n+\hat{1},1)$ (the one whose corners are $n_{1,6,5,4}$ and $n'_{1,6,5,4}$). We define the Wilson line
\begin{equation}
    T_n(x)\equiv (00''a''e''xb''1''1)
\end{equation}
which runs from $n_1=n+\hat{1}$ to $n_0=n$. Notice that when $x\notin \partial c(n+\hat{1},1)$, the $(e''x)$ segment runs in the \emph{interior} of the hypercube, not in any cube on the boundary of the hypercube, so it seems our interpolation introduced in the previous subsection becomes insufficient. However, as we will see, our final result will only involve $x\in \partial c(n+\hat{1},1)$, for which case all segments of $T$ runs in the cubes on the boundary $\partial h(n)$. More exactly, when $x$ lies on one of those plaquettes on $\partial c(n+\hat{1},1)$, we may check
\begin{equation}
\begin{aligned}
    T_n(x)&=R_{n,\mu;1}(x) \ \ \ \mbox{for } x\in p(n+\hat 1,1,\mu), \\
    T_n(x)&=R_{n,1;\mu}(x-\hat{1})R_{n+\hat \mu,\mu;1}(x)R_{n+\hat 1,1;\mu}(x)^{-1} \ \ \ \mbox{for } x\in p(n+\hat 1+\hat \mu,1,\mu).
\end{aligned}
\label{eqn:T}
\end{equation}
So our final result will only involve the $R$ functions which are expressed by the $g$ and $h$ interpolation functions that have already been constructed.

The expression for the CS phase saddle around the hypercube $h(n)$ is (see Appendix \ref{app:q_density} for derivation)
\begin{equation}
\begin{aligned}
    & \ (\mathrm{d}\alpha^{(0)})_h \mod 2\pi\\[.2cm]
    =& -\frac{\epsilon^{1\mu\nu\rho}}{12\pi}\int_{c(n+\hat 1,1)}\mathrm{d}^3 x \: \Tr\left[T_n(x)^{-1}\partial_\mu T_n(x) T_n(x)^{-1}\partial_\nu T_n(x) T_n(x)^{-1}\partial_\rho T_n(x)\right]\\
    & -\frac{\epsilon^{1\mu\nu\rho}}{4\pi}\int_{p(n+\hat 1+\hat \mu,1,\mu)}\mathrm{d}^2 x \: \Tr\left[ T_n(x)^{-1} \left( R_{n,1;\mu}(x-\hat{1})  \partial_\nu R_{n+\hat \mu,\mu;1}(x) \partial_\rho R_{n+\hat 1,1;\mu}(x)^{-1} \right. \right. \\
    & \phantom{\int} \left.\left.  \hspace{6cm} + \partial_\nu R_{n,1;\mu}(x-\hat{1})  R_{n+\hat \mu,\mu;1}(x) \partial_\rho R_{n+\hat 1,1;\mu}(x)^{-1} \right. \right. \\
    & \phantom{\int} \left.\left. \hspace{6cm} + \partial_\nu R_{n,1;\mu}(x-\hat{1}) \partial_\rho R_{n+\hat \mu,\mu;1}(x) R_{n+\hat 1,1;\mu}(x)^{-1} \right) \right] \\
    & \ \ \mod 2\pi \ .
\end{aligned}
\label{eqn:q_density}
\end{equation}
Naively, the first line after the equality seems to involve $x$ in the interior of the cube $c(n+\hat{1},1)$, which will in turn make $T_n(x)$ involve gauge field interpolation into the interior of the hypercube $h(n)$ which is undesired (recall the discussion above (\ref{eqn:T})). However, we note this is a Wess-Zumino-Witten (WZW) integral, whose result actually only depends on $T_n(x)$ on the boundary $x\in \partial c(n+\hat{1},1)$ given by (\ref{eqn:T}), up to a $2\pi\mathbb{Z}$ which we will drop anyways (since it can be absorbed into the dynamical integer part $s_h$ of the instanton density $q_h$). Topologically, $T_n(x)$ with $x\in \partial c(n+\hat{1},1)$ is mapping a topological 2d sphere into $SU(2)$ which is a 3d sphere, and this WZW integral is evaluating the 3d volume bounded by the embedded 2d sphere, under the normalization that the volume of the entire $SU(2)$ is $2\pi$.
\footnote{The relation (\ref{eqn:q_density}) between the CS on the boundary of a 4d region and the WZW constructed out of Wilson lines with one end on the boundary of a 3d region is a reminiscence of the categorical delooping procedure, braiding data and Yang-Baxter equation constraint discussed in \cite{Chen:2024ddr}. Here we will not plunge into the depth.}
This topological picture also justifies an implicit but crucial assumption here, that having defined $T_n(x)$ on $x\in \partial c(n+\hat{1},1)$ via (\ref{eqn:T}), there indeed exists \emph{some} smooth definition of $T_n(x)$ for all $x\in c(n+\hat{1},1)$---this is because a 2d sphere embedded in a 3d sphere can always be contracted.

If we want an integral that manifestly involves $x\in\partial c(n+\hat{1},1)$ only, we may diagonalize
\begin{equation}
    T_n(x)=U(x) e^{i\Lambda(x)} U(x)^{-1}.
    \label{eqn:diag_T_app}
\end{equation}
Then the WZW integral can be written as
\begin{equation}
\begin{aligned}
    &\int_{c}\Tr\left(T_n^{-1}\mathrm{d}T_n\wedge T_n^{-1}\mathrm{d}T_n\wedge T_n^{-1}\mathrm{d}T_n\right)\\
    &=\int_{\partial c}\Tr\left[6\left(i\mathrm{d}\Lambda \wedge U^{-1}\mathrm{d}U \right)-3\left(e^{i\Lambda}U^{-1}\mathrm{d}U\wedge e^{-i\Lambda}U^{-1}\mathrm{d}U \right)\right]
\end{aligned}
\label{eqn:T_int}  
\end{equation}
where the 2d integrand is known as the WZW curving. Note that in the diagonalization process:
\begin{itemize}
\item $U$ and $\Lambda$ are not unique, since we can act on the right of $U$ a unitary that commutes with $e^{i\Lambda}$, and also $\Lambda$ can be shifted by $2\pi\mathbb{Z}$ diagonal matrices. One can show such non-uniqueness can at most change the result by $2\pi \mathbb{Z}$,
\footnote{This statement is part of the proof that the WZW integral over a closed 3d space is an integer, see e.g. \cite{Chen:2024ddr,Gawedzki:2002se}.} 
which we are dropping anyways.
\item Though non-unique, the choice $U(x)$ and $\Lambda(x)$ can be made smoothly over $x\in\partial c(n+\hat 1,1)$, because $T_n(x)$ is smoothly defined for the cube $c(n+\hat 1,1)\ni x$ (not just $\partial c(n+\hat 1,1)$), and a cube is contractible.
\end{itemize}
Then we have an expression that is manifestly determined by the $g$ and $h$ interpolations.

There are two crucial properties of (\ref{eqn:q_density}). First, the expression (\ref{eqn:q_density}) is manifestly gauge invariant. By definition, $R_{m,\mu;\nu}$ is a Wilson line from $m+\hat \nu$ to $m$. Since our interpolation protocol is gauge covariant, under a gauge transformation $w$, the $R$ function transforms as
\begin{equation}
    R_{m,\mu;\nu}(x)\rightarrow w(m)R_{m,\mu;\nu}(x)w(m+\hat \nu)^{-1} .
    \label{eqn:R_gauge_transf}
\end{equation}
Crucially, $w$ does not depend on $x$ when we take $x$-derivative of $R$. Therefore, substituting the transformed $R$'s into Eq.(\ref{eqn:q_density}), obviously the constructed CS phase saddle is gauge invariant. Moreover, at any intermediate step, there is no involvement of extra gauge choice except for those at the lattice vertices (\ref{eqn:R_gauge_transf}), since we only used Wilson lines that start and end at lattice vertices (this is in contrast to \cite{Luscher:1981zq}, see Appendix \ref{app:q_density}).

Second, if (\ref{eqn:q_density}) indeed is some lattice exterior derivative, then it should be true that
\begin{equation}
    \sum_h (\mathrm{d}\alpha^{(0)})_h\mod 2\pi = 0 \mod 2\pi
    \label{eqn:sum_q_int}
\end{equation}
when the 4d Euclidean manifold where the lattice is embedded is closed. This is most easily seen from the original expression in \cite{Luscher:1981zq}, see the end of Appendix \ref{app:q_density}.

In summary, the expression for the CS phase saddle around a hypercube only depends on the $R$ function, which is by definition some Wilson line connecting vertices of the lattice, and whose explicit expression is given by the interpolation functions $g$ and $h$. So we can explicitly compute the CS phase saddle $e^{i(\mathrm{d}\alpha^{(0)})_h}$ given all the link and plaquette dynamical variables sampled on the boundary of the hypercube.

\subsection{The CS Sensitivity $|\nu_c|$}
\label{ss_CS_sensitivity}

As introduced in Section \ref{s_path_int}, the CS sensitivity $|\nu|$ controls how strongly the dynamical CS phase $e^{i\alpha_c}$ fluctuates around the interpolated CS phase saddle $e^{i\alpha^{(0)}_c}$, or equivalently, how strongly $e^{i\beta_c}$ fluctuates around $1$. We have explained its role in Section \ref{s_path_int} that when we run into singularity in the interpolations in Section \ref{ss_interpolation}, this sensitivity should be approach zero. And in \ref{sec:interpolation_cube}, we have briefly discussed the cases when there will be singularities of the $h$ functions. In this subsection, we summarize these requirements on the $|\nu_c|$ function, which will also suggest in which variables $|\nu_c|$ should depend on. Since the possible singularity originates from the $h_{\xi_1\xi_2}$ functions, we will present the requirements separately for different values of $\xi_1$ and $\xi_2$ (recall \ref{sec:interpolation_cube}).

\vspace{.3cm}\noindent
$\boldsymbol{\xi_1=\xi_2=+.}$ In this case, $h$ function is singular if there exists a $y^\gamma\in [0,1]$ such that $u_c(y^\gamma)=-1$. Thus we require that
\begin{equation}
    |\nu_c|_{++}=0 \text{ if } \min\left\{\Tr \, u_c(y^\gamma)\middle| y^\gamma\in[0,1]\right\}=-2.
\end{equation}
Note that we have used $\Tr \, u_c$ since we want our construction to be gauge invariant. By this requirement we can see that the minimal reasonable choice of $|\nu_c|_{++}$ is to make it an increasing function of $\min\left\{\Tr \, u_c(y^\gamma)\middle| y^\gamma\in[0,1]\right\}$, such that when the argument decreases from $2$ to $-2$, $|\nu_c|_{++}$ decreases from $|\nu|^{\text{max}}_{++}=1$ to $0$ (but $W_3$ is always positive). 

Apart from that, $|\nu_c|_{++}$ may also depend on other \emph{gauge invariant} variables, such as the entire profile of $u_c(y^\gamma)$ for $0\leq y^\gamma \leq 1$. The same is understood in the rest of the cases below.

\vspace{.3cm}\noindent
$\boldsymbol{\xi_1=\xi_2=-.}$ In this case, as we have briefly discussed in \ref{sec:interpolation_cube}, there are two types of singularities. Similar to the previous case, we require that
\begin{equation}
    |\nu_c|_{--}=0 \text{ if } \max\left\{\Tr \, u_c(y^\gamma)\middle| y^\gamma\in[0,1]\right\}=2 \text{ \emph{or} } \Tr\left[(\hat{k}_1\cdot \vec\sigma)(03)(\hat{k}_2\cdot\vec\sigma)(30)\right]=-2.
\end{equation}
So the minimal reasonable choice of $|\nu_c|_{--}$ is that it decreases from some $0<|\nu|^{\text{max}}_{--}<1$ to $0$ as $\max\left\{\Tr \, u_c(y^\gamma)\middle| y^\gamma\in[0,1]\right\}$ increases from $-2$ to 2 \emph{or} $\Tr\left[(\hat{k}_1\cdot \vec\sigma)(03)(\hat{k}_2\cdot\vec\sigma)(30)\right]$ decreases from 2 to $-2$.

\vspace{.3cm}\noindent
$\boldsymbol{\xi_1=-,\xi_2=+.}$ In this case the $h$ function is more complicated thus resulting in more singularities. We require that
\begin{equation}
\begin{aligned}
    |\nu_c|_{-+}=0 \text{ if }& \max\left\{\Tr \, u_c(y^\gamma)\middle| y^\gamma\in[0,1]\right\}=2,  \min\left\{\Tr \, u_c(y^\gamma)\middle| y^\gamma\in[1/2,1]\right\}=-2\\
    &\text{ \emph{or} } \Tr\left[(\hat t_c^{(1/2)}\cdot \vec\sigma)(\hat{k}_1\cdot \vec\sigma)\right]=-2.
\end{aligned}
\end{equation}
The minimal reasonable choice of $|\nu_c|_{-+}$ is that it decreases from some $0<|\nu|^{\text{max}}_{-+}<1$ to $0$ as $\max\left\{\Tr \, u_c(y^\gamma)\middle| y^\gamma\in[0,1]\right\}$ increases from $-2$ to 2, $\min\left\{\Tr \, u_c(y^\gamma)\middle| y^\gamma\in[1/2,1]\right\}$ decreases from 2 to $-2$ \emph{or} $\Tr\left[(\hat t_c^{(1/2)}\cdot \vec\sigma)(\hat{k}_1\cdot \vec\sigma)\right]$ decreases from 2 to $-2$.

\vspace{.3cm}\noindent
$\boldsymbol{\xi_1=+,\xi_2=_.}$ This case is similar to previous case and we require that
\begin{equation}
\begin{aligned}
    |\nu_c|_{+-}=0 \text{ if }& \max\left\{\Tr \, u_c(y^\gamma)\middle| y^\gamma\in[0,1]\right\}=2,  \min\left\{\Tr \, u_c(y^\gamma)\middle| y^\gamma\in[0,1/2]\right\}=-2\\
    &\text{ \emph{or} } \Tr\left[(\hat t_c^{(1/2)}\cdot \vec\sigma)(03)(\hat{k}_2\cdot \vec\sigma)(30)\right]=-2.
\end{aligned}
\end{equation}
The minimal reasonable choice of $|\nu_c|_{-+}$ is that it decreases from $|\nu|^{\text{max}}_{+-}=|\nu|^{\text{max}}_{-+}$ to 0 as $\max\left\{\Tr \, u_c(y^\gamma)\middle| y^\gamma\in[0,1]\right\}$ increases from $-2$ to 2, $\min\left\{\Tr \, u_c(y^\gamma)\middle| y^\gamma\in[0,1/2]\right\}$ decreases from 2 to $-2$ \emph{or} $\Tr\left[(\hat t_c^{(1/2)}\cdot \vec\sigma)(03)(\hat{k}_2\cdot \vec\sigma)(30)\right]$ decreases from 2 to $-2$.

\vspace{.3cm}
To conclude, we have listed the requirements of $|\nu_c|$ in all four cases. In these requirements, we have introduced some variables that $|\nu_c|$ must depend on and have briefly sketched its minimal choices. $|\nu_c|$ can also depend on other \emph{gauge invariant} variables. In the actual Monte-Carlo calculation, the specific expression of $|\nu_c|$ should be determined through optimization of numerical behavior.

\section{Generalization to $SU(N)$}
\label{s_SUN}

The generalization from $SU(2)$ to $SU(N)$ has been briefly sketched in \cite{Chen:2024ddr}.
\footnote{The group theoretic nature of the construction originated from \cite{Gawedzki:2002se}; for generalizations towards more general Lie groups, consult \cite{Gawedzki:2003pm}. \ However, note that in \cite{Gawedzki:2002se} there is only the counterpart of what we called $\xi$ but no counterpart of $\hat{k}$. This is because we only need the latter for gauge covariance purpose---recall Eq.(\ref{eqn:k_gauge_transf})---which took no place in \cite{Gawedzki:2002se}, but were it for purely topological purpose, one could have simply fixed some $\hat{k}$ direction.}
Here, for this work to be self-contained, we present the generalization in more explicit details. The main task is to generalize Fig.~\ref{fig:g_sketch}.

\begin{figure}
    \centering
    \includegraphics[width=0.7\linewidth]{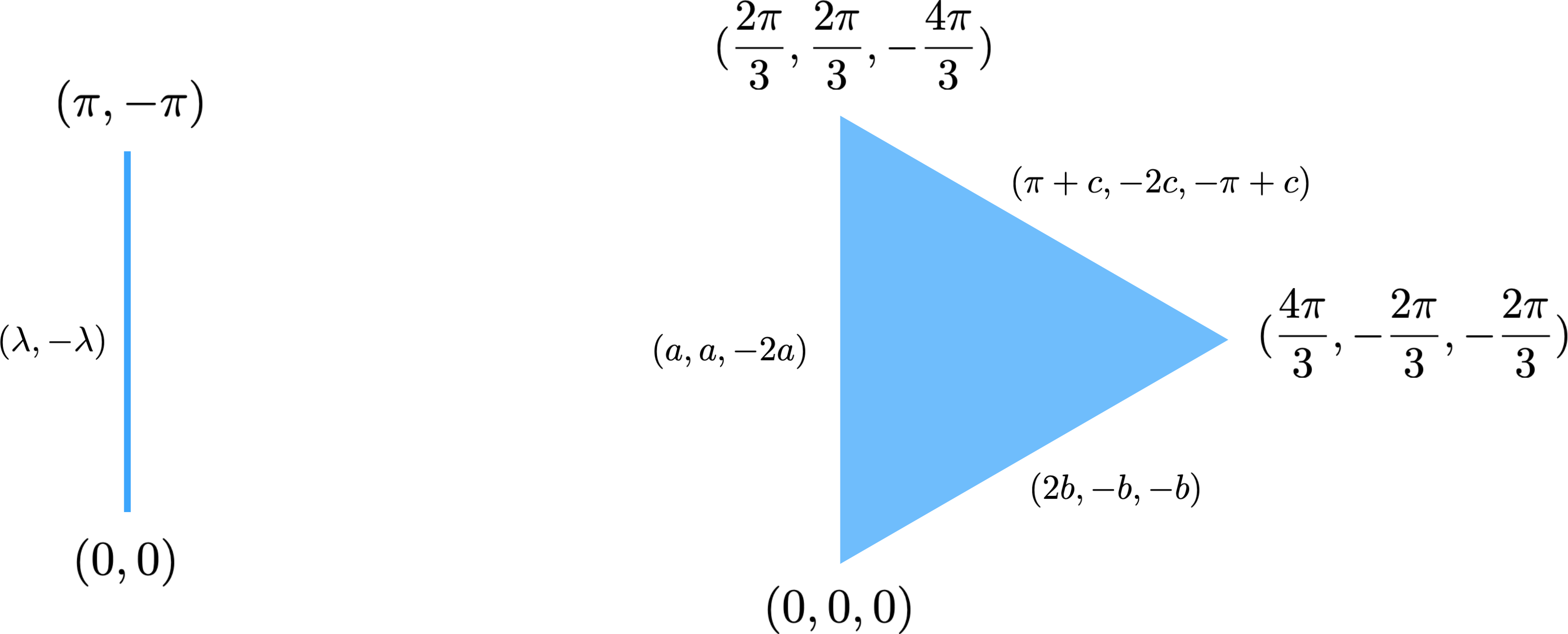}
    \caption{The Weyl alcove for $SU(2)$ and $SU(3)$ respectively. The numbers (after multiplying by $i$) are the eigenvalues in the Cartan subalgebra.}
    \label{fig:Weyl_Alcove}
\end{figure}

Consider the Cartan subalgebra of the Lie algebra $\mathfrak{su}(N)$, which is the $(N-1)$-dimensional linear space of $N\times N$ diagonal traceless Hermitian matrices, thought of as the eigenvalues of an $\mathfrak{su}(N)$ matrix (up to a conventional $i$ factor); inner product is given by the Killing form $\propto\Tr(\cdot, \cdot)$. But permutations of eigenvalues (action of Weyl group) is unimportant, so we may for instance demand the eigenvalues to be ordered in a non-increasing manner, and such a region, obtained by modding out the Weyl group in the Cartan subalgebra, is called a Weyl chamber. A Weyl alcove is a region in a Weyl chamber with not too large eigenvalues, so that when exponentiating $\mathfrak{su}(N)$ to $SU(N)$, the exponentiation of the Weyl alcove will cover the space of possible eigenvalues of $SU(N)$ in a 1-to-1 manner. See examples for $SU(2)$ and $SU(3)$ in Fig.~\ref{fig:Weyl_Alcove}. In general, the Weyl alcove of $SU(N)$ is an $(N-1)$-simplex, cornered at $\mathfrak{su}(N)$ eigenvalues
\begin{equation}
\xi^z=\left(\underbrace{\frac{2\pi z}{N},\dots,\frac{2\pi z}{N}}_{N- z},\underbrace{-\frac{2\pi(N- z)}{N},\dots,-\frac{2\pi(N- z)}{N}}_ z\right),\quad  z = 0, 1,2,\dots,N-1
\label{eqn:Weyl_Alcove_corners}
\end{equation}
(so the simplex is not equilateral for $N>3$). The $(N-2)$-dimensional faces on the boundary of the simplex exponentiate to $SU(N)$ matrices with $2$-fold degenerate eigenvalues (i.e. with only $N-1$ distinct eigenvalues), and $(N-m)$-dimensional faces on the boundary of the simplex exponentiate to $SU(N)$ matrices with only $N-m+1$ distinct eigenvalues; in particular the $N$ 0-dimensional corners \eqref{eqn:Weyl_Alcove_corners} of the simplex exponentiate to the $Z(SU(N))\cong\mathbb{Z}_N$ center of $SU(N)$.

To generalize Fig.~\ref{fig:g_sketch} from $SU(2)$ to $SU(N)$, let the holonomy $u_p\in SU(N)$ around a plaquette be diagonalized as $u_p=\mathcal{U}_p e^{i\mathcal{D}_p} \mathcal{U}_p^{-1}$ where $i{\mathcal{D}}_p$ belongs to the Weyl alcove; $\mathcal{U}_p$ is only well defined up to $\mathcal{U}_p \rightarrow \mathcal{U}_p \mathcal{V}_p$ where $\mathcal{V}_p$ commutes with $e^{i\mathcal{D}_p}$. We need to consider interpolations from $1$ to $u_p$. The ``most direct" way is to interpolate from the origin $0$ to $\mathcal{D}_p$ in the Weyl alcove along the straight path, so that the interpolated path in $SU(N)$ is
\begin{align}
\{ \mathcal{U}_p e^{iy\mathcal{D}_p} \mathcal{U}_p^{-1} | 0\leq y\leq 1 \} \ .
\end{align}
However, this interpolation becomes ambiguous when $\mathcal{D}_p$ approaches the face of the alcove opposite to the origin, because on that face, the largest and the smallest entries in $\mathcal{D}_p$ differ by $2\pi$ and are therefore degenerate when exponentiated; $\mathcal{U}_p$ therefore has a $U(2)$ ambiguity $\mathcal{V}_p$ that rotates these two eigenvalues in $e^{i\mathcal{D}_p}$. But for $0<y<1$, $e^{iy\mathcal{D}_p}$ does not commute with such $\mathcal{V}_p$, and hence the interpolation path becomes ambiguous. Therefore we need other ways of interpolation. Obviously, we can use $N$ ways of interpolation, labeled by $\xi_p\in \{\text{corners of the Weyl alcove}\}\cong \mathbb{Z}_N$, see \eqref{eqn:Weyl_Alcove_corners}; for $SU(2)$, $\xi_p=\xi^0=0$ and $\xi_p=\xi^1=(\pi,-\pi)$ were respectively labeled by $\xi_p=\pm$ before. Note this does not mean $\xi_p$ admits any meaningful composition, so they are just labels but do not form a $\mathbb{Z}_N$ group. For $\xi_p=0$, the interpolation path is described above, while for other $\xi_p$, the interpolation path consists of two segments:
\begin{align}
    \{ \mathcal{K}_p e^{i2y \xi_p} \mathcal{K}_p^{-1} | 0\leq y\leq 1/2 \} \ \cup \ \{ \mathcal{U}_p e^{i(2-2y)\xi_p+i(2y-1)\mathcal{D}_p} \mathcal{U}_p^{-1} | 1/2\leq y\leq 1 \}
\end{align}
where $\mathcal{K}_p \in SU(N)/G_{\xi_p}$ with the $G_{\xi_p}$ denoting the matrices that commute with $\xi_p$ (and hence with any $e^{i2y \xi_p}$), and from \eqref{eqn:Weyl_Alcove_corners} we can see $G_{\xi^ z}=U( z)\times U(N- z)/U(1)$. For $SU(2)$, $G_{\xi_p=-}=U(1)\times U(1)/U(1)\cong U(1)$ and the $\mathcal{K}_p \in SU(2)/U(1)$ was labelled by $\hat{k}_p\in S^2$ before. Both $\xi_p$ and $\mathcal{K}_p$ are to be dynamically sampled in the path integral. And under a gauge transformation we have $u_p\rightarrow w_p u_p w_p^{-1}$, thus we expect the matrix $\mathcal{K}_p$ to transform as
\begin{equation}
    \mathcal{K}_p\rightarrow [w_p \mathcal{K}_p]\in SU(N)/G_{\xi_p}
\end{equation}
so that the path integral is gauge invariant.

The plaquette weight $W_2$ now has dependence $W_2(\mathcal{D}_p, \xi_p)$, which should be invariant under the symmetries of the Weyl alcove fixing the origin, i.e. reversing the sign of the $\mathfrak{su}(N)$ eigenvalues. For given $\xi_p$, $W_2$ should decrease with the distance between $\mathcal{D}_p$ and $\xi_p$ in the Weyl alcove; as $\mathcal{D}_p$ approaches the face of the alcove that is opposite to $\xi_p$, $W_2$ should decreases towards $0$, just like in Fig.~\ref{fig:W2_idea}, because there the second segment of the interpolation becomes ambiguous due to the ambiguity in $\mathcal{U}_p$. Moreover, when $D_p=\xi_p$, we require $W_2(0, 0)$ to be large, while all other $W_2(\xi_p, \xi_p)$ to be small (and respect the symmetry of the Weyl alcove of reversing the sign of the $\mathfrak{su}(N)$ eigenvalues), for the same physical intuition as in Fig.~\ref{fig:W2_idea}. The cube weight $W_3$, whose $\nu$ function depends on $u_p, \xi_p, \mathcal{K}_p$ around the cube, can be technically constructed along the same line as in Section \ref{s_CS}.

\section{Further Discussions}
\label{s_remarks}

In this work we introduced an explicit categorical refinement (\ref{eqn:path_int_convenient}) of lattice Yang-Mills theory in 4d (and higher dimensions), with the main technical step, i.e. the construction of $e^{id\alpha_h^{(0)}}$ and $|\nu_c|$, explained in Section \ref{s_CS}. Our construction is explicit in the sense that the concept, the degrees of freedom and the key function $d\alpha_h^{(0)}$ are explicit given, while the detailed weight functions $W_2, W_3, W_4$ and $|\nu_c|$ are subjected to some explicitly stated requirements, but otherwise their detailed forms are to be optimized in numerical practices---just like the traditional plaquette weight $W_2^{(0)}$ in Wilson's construction. Our categorical refinement allows the instanton density $q_h$ over a hypercube $h$ to be naturally defined. Moreover, in 3d, the categorical refinement allows the lattice CS term to be defined.

\

\noindent\textbf{Numerical Explorations}

The primary application in mind, apparently, is so that generic correlation functions that involve instanton number or instanton density can now be unambiguously computed in Monte-Carlo. Moreover, now that we have an explicit local lattice definition of instanton density regardless of whether the instanton is ``large" or ``small", it will be interesting and practically important to understand the influence on the topological freezing problem \cite{Luscher:2010iy, Schaefer:2010hu}, i.e. what such a local definition might do to the tunneling of instanton number in local Monte-Carlo updates.

More generally, we should explore the influence of our refinement on the lattice renormalization flow. Recall the Villainization of $S^1$ non-linear sigma model \cite{Berezinsky:1970fr, Villain:1974ir} facilitated the analysis of the renormalization towards BKT transition by signifying the running of the vortex fugacity \cite{jose1977renormalization}---a plaquette weight that is not present in the traditional $S^1$ non-linear sigma model (XY model). The idea is that, even if one does not include a vortex fugacity weight to begin with, such weight will be dynamically generated as the lattice is coarse-grained, so we should just explicitly introduce such weight to begin with in order to keep track of the associated effects. Now, our refinement of lattice Yang-Mills theory is just a categorical generalization of Villainization \cite{Chen:2024ddr}, with the same goal to better conenct the lattice QFT (which is well-defined) to continuum QFT (which is desired but not well-defined), so it is reasonable to anticipate that the new degrees of freedom and the new weights that we introduced on the higher dimensional cells may facilitate the lattice renormalization in a similar fashion. But this can only be explored numerically in the strong coupling regime.

At a technical level, we can note that, with non-trivial cube weight $W_3$ and hypercube weight $W_4$, integrating out the new degrees of freedom will introduce effective couplings between the $u_p$ on neighboring plaquettes (as opposed to merely (\ref{eqn:W2_recovery}) which is only applicable when $W_3=1$, $W_4=1$). At least at a conceptual level, the Symanzik improvement \cite{Symanzik:1983dc, Weisz:1982zw, Curci:1983an} is indeed to directly introduce such couplings in order to improve the lattice renormalization. Now, such effective couplings will be dynamically generated even if we did not directly introduce them, and moreover $W_3, W_4$ from which they are generated are to be optimized by numerical performance and have intuitive physical meanings. This is the argument from another perspective that the new degrees of freedom and new weight may help with the lattice renormalization.

\

\noindent\textbf{Analytical Explorations}

There are many important formal, mathematical problems that arises from the categorical perspective of lattice QFT, see \cite{Chen:2024ddr}. In this paper, rather, we will focus on those problems that are directly related to having an explicit construction.

While in our construction the detailed weight functions should, ideally, be optimized numerically, we may as well just propose some explicit form of the weight functions, and ask whether in certain limits of parameters some physical observables such as the topological suspectibility $\propto \langle Q^2 \rangle$ can be calculated perturbatively. For 3d Chern-Simons-Yang-Mills, this task becomes even more prominent, because the CS theory cannot be simulated by sign free Monte-Carlo. While a full solution like that for Chern-Simons-Maxwell \cite{Xu:2024hyo} is impossible, one shall still explore whether in certain limit the properties of this lattice theory can be perturbatively computed and matches with those computations in the continuum \cite{Pisarski:1985yj}.
\footnote{We are grateful to Robert Pisarski for suggesting these problems.}

Another direction is to couple the refined pure lattice Yang-Mills theory to other fields, in the purview of symmetries and anomalies. There are two obvious tasks in this direction:
\begin{enumerate}
    \item A categorically refined $|SU(N)|$ lattice non-linear sigma model has also been proposed in \cite{Chen:2024ddr}, whose explicitly well-defined skrymions are interpreted as baryons. Coupling this theory to the categorically the refined $SU(N)$ lattice gauge theory, we expect the instanton fluctuation to explicitly break the $U(1)$ baryon conservation symmetry of the non-linear sigma model. This would be a lattice manifestation of the mixed anomaly between the $SU(N)$ global symmetry (before gauging) of non-linear sigma model and the $U(1)$ baryon conservation symmetry.
    \item The pure $SU(N)$ Yang-Mills theory itself has a 1-form $\mathbb{Z}_N$ global symmetry, seen by $u_l\rightarrow e^{i2\pi n_l/N} u_l$ where $n_l\in\mathbb{Z}_N$ and $dn_p=0\mod N$ so that $u_p$ remains invariant. This 1-form global symmetry has a mixed anomaly with the $2\pi$ periodicity of the topological theta term, and this should be manifested by coupling the theory to a 2-form $\mathbb{Z}_N$ background gauge field. This has been demonstrated on the lattice using L\"uscher's geometrical construction \cite{Abe:2023ncy}, and should be repeated in the full categorical construction as well. Related to this, we should demonstrate that the 1-form $\mathbb{Z}_N$ global symmetry becomes self-anomalous when a non-trivial CS level in present in 3d.
\end{enumerate}
These are kinematics problems that do not require an analysis of the dynamics.

Finally, a highly ambitious long term problem is in the context of \emph{constructive QFT}---the programme to make sense of continuum QFT, for which one major approach is to understand the continuum limit of lattice QFT. Remarkable analysis has been performed by Balaban on Wilson's traditional lattice gauge theory, and partial result has been achieved highly non-trivially \cite{Balaban:1985pi,Balaban:1989ay,Balaban:1989qz}. One can ask how the refinement, which is supposed to bring the lattice theory closer to the desired continuum theory (at least at the topological level), might alter the analysis.

\vspace{1cm} 

\noindent \emph{Acknowledgement.} We thank Robert Pisarski for suggesting ideas for future analytical explorations. This work is supported by NSFC under Grants No.~12447104, No.~12174213 and No.~12342501.

\vspace{.5cm}

\appendix

\section{Derivation of Eq.(\ref{eqn:q_density}) in Relation to L\"{u}scher's Geometric Construction}
\label{app:q_density}

\begin{figure}
    \centering
    \includegraphics[width=0.4\linewidth]{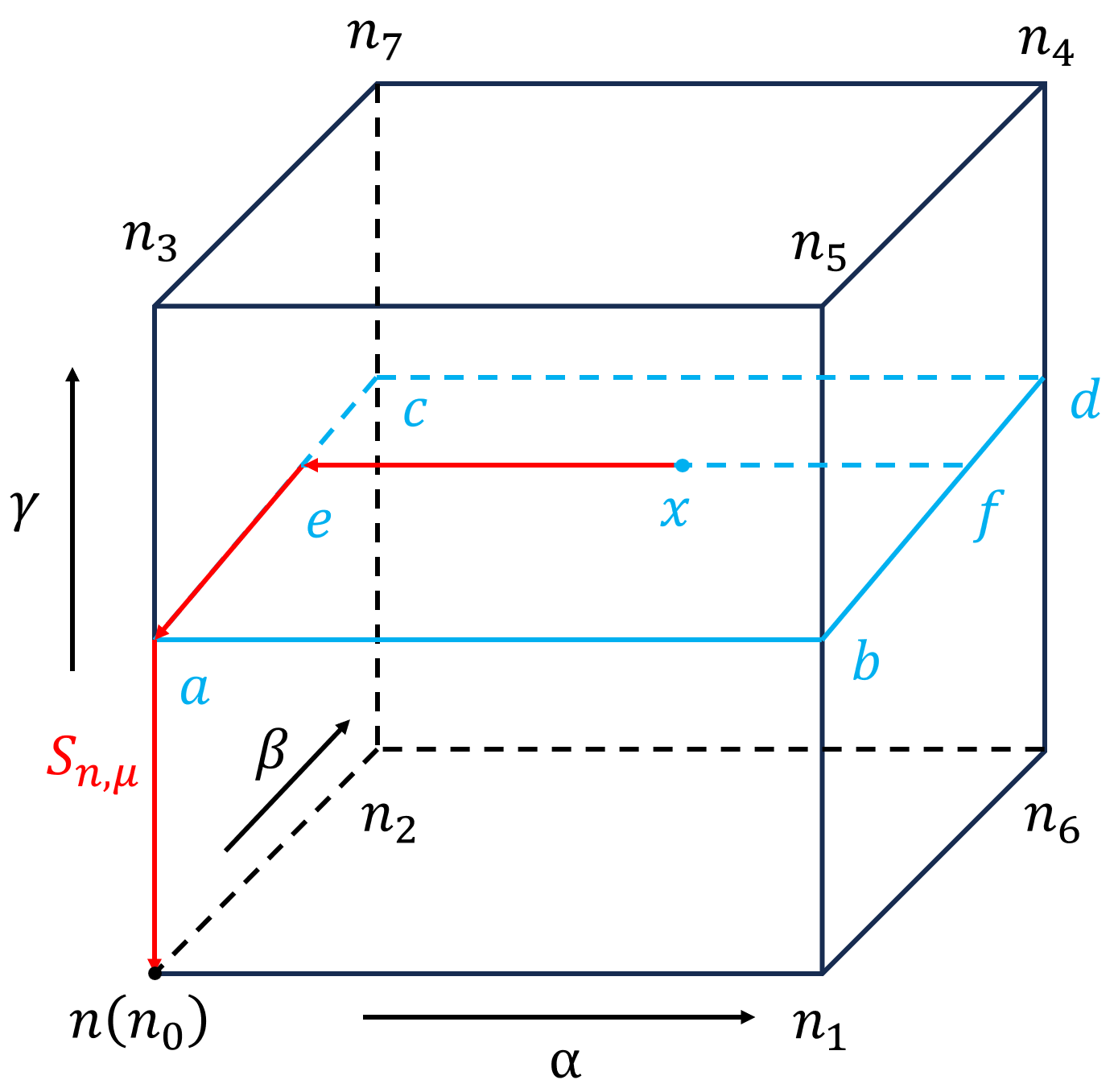}
    \caption{Definition of the Wilson line $S_{n,\mu}(x)$ in the cube $c(n,\mu) \ (n=n_0)$. The coordinate directions $\{\alpha,\beta,\gamma\}=\{1,2,3,4\}\backslash \{\mu\}$ and $\alpha<\beta<\gamma$. $S_{n,\mu}$ is the red Wilson line from $x$ to $n$.}
    \label{fig:S}
\end{figure}

\begin{figure}
    \centering
    \includegraphics[width=0.3\linewidth]{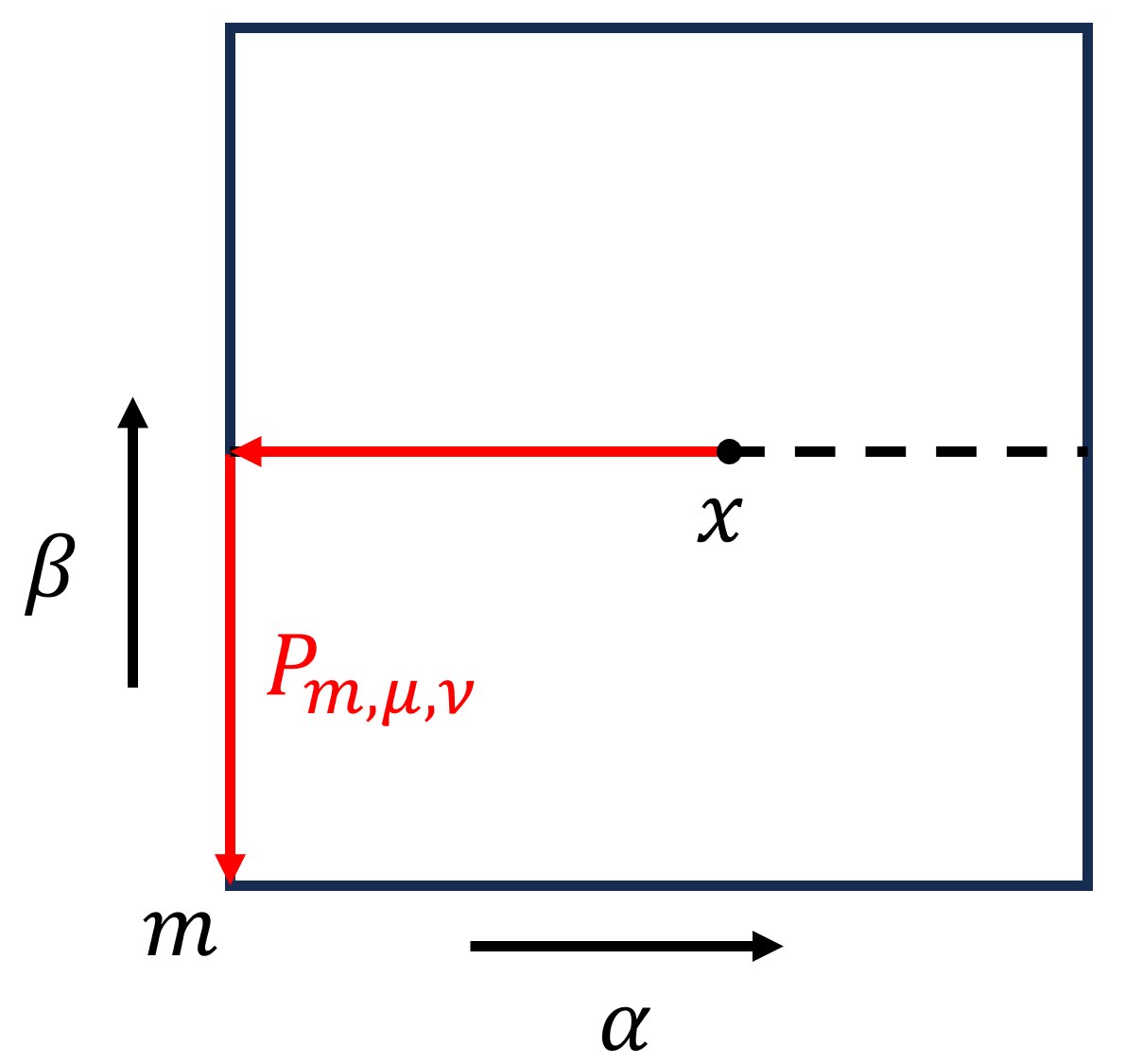}
    \caption{Definition of $P_{m,\mu,\nu}(x)$ for $x\in p(m,\mu,\nu)$. $P_{m,\mu,\nu}(x)$ is the red Wilson line from $x$ to $m$. The coordinate directions $\{\alpha,\beta\}=\{1,2,3,4\}\backslash \{\mu,\nu\}$ and $\alpha<\beta$.}
    \label{fig:P_def}
\end{figure}

First we briefly review the result in \cite{Luscher:1981zq}. We need to introduce some Wilson lines. For $x\in c(n=n_0,\mu)$, define $S_{n,\mu}(x)=(0aex)$ as the red Wilson line from $x$ to $n=n_0$ in Fig.~\ref{fig:S}.
And for $x\in p(m,\mu,\nu)$, define $P_{m,\mu,\nu}(x)$ as the red Wilson line from $x$ to $m$ in Fig.~\ref{fig:P_def}.
So for $x\in p(n,\mu,\nu)$ we have
\begin{equation}
    S_{n,\mu}(x)=P_{n,\mu,\nu}(x),
\end{equation}
and for $x \in p(n+\hat\lambda,\mu,\lambda)$ we have,
\begin{equation}
     S_{n,\mu}(x)=R_{n,\mu;\lambda}(x)P_{n+\hat{\lambda},\mu,\lambda}(x)
\end{equation}
which explains the motivation of the $R$ functions in (\ref{eqn:R_def}). Note that specifying $S$ and $P$ would essentially specify some gauge choice in the interior of the cube and the plaquette, since they are not closed Wilson loops that starts and end on vertices.

It was derived in \cite{Luscher:1981zq} that around the hypercube $h(n)$ one can define
\begin{equation}
\begin{aligned}
    &(\mathrm{d}\alpha^{(0)})_{h(n)}\\
    &=\frac{\epsilon^{\mu\nu\rho\sigma}}{12\pi}\left\{\int_{c(n+\hat{\mu},\mu)}\mathrm{d}^3x \: \Tr[(S_{n+\hat{\mu},\mu})^{-1}\partial_\nu S_{n+\hat{\mu},\mu}(S_{n+\hat{\mu},\mu})^{-1}\partial_\rho S_{n+\hat{\mu},\mu}(S_{n+\hat{\mu},\mu})^{-1}\partial_\sigma S_{n+\hat{\mu},\mu}]\right.\\
    &\quad+3\int_{p(n+\hat{\mu}+\hat{\nu},\mu,\nu)}\mathrm{d}^2x \: \Tr[P_{n+\hat{\mu}+\hat{\nu},\mu,\nu}\partial_\rho(P_{n+\hat{\mu}+\hat{\nu},\mu,\nu})^{-1} (R_{n+\hat{\mu},\mu;\nu})^{-1}\partial_\sigma R_{n+\hat{\mu},\mu;\nu}]\\
    &\quad-\int_{c(n,\mu)}\mathrm{d}^3x \: \Tr[(S_{n,\mu})^{-1}\partial_\nu S_{n,\mu}(S_{n,\mu})^{-1}\partial_\rho S_{n,\mu}(S_{n,\mu})^{-1}\partial_\sigma S_{n,\mu}]\\
    &\quad\left.-3\int_{p(n+\hat{\nu},\mu,\nu)}\mathrm{d}^2x \: \Tr[P_{n+\hat{\nu},\mu,\nu}\partial_\rho(P_{n+\hat{\nu},\mu,\nu})^{-1} (R_{n,\mu;\nu})^{-1}\partial_\sigma R_{n,\mu;\nu}]\right\}.
    \label{eqn:Luscher_q_density}
\end{aligned}
\end{equation}
In \cite{Luscher:1981zq}, this was recognized as $2\pi$ times the instanton density over $h(n)$. However, by specifying the Wilson lines $S$ which are only interpolated into the interior of the cubes on the boundary of the hypercube $h$, without specifying any interpolation into the interior of the hypercube, we can readily anticipate (recall the discussion below (\ref{eqn:q_CS_cont})) that the integer part of the expression above must be gauge dependent. In the below we will soon see this is indeed the case.

We want to turn Eq.(\ref{eqn:Luscher_q_density}) from \cite{Luscher:1981zq} into such a form that manifests the fact that its $\mod 2\pi$ part is gauge independent, and discard its gauge dependent $2\pi\mathbb{Z}$ part. The reasons being:
\begin{itemize}
\item Recall the path integral (\ref{eqn:path_int_convenient}). By construction, we only need the $\mod 2\pi$ part of $(\mathrm{d}\alpha^{(0)})_{h(n)}$ anyways, to be interpreted as the total CS phase saddle on the cubes around the hypercube $h(n)$. The integer part of the instanton density over $h(n)$ is a dynamical variable $s_h$.

\item Avoiding gauge choice in the expression not only ``makes things nice'', it is practically important for avoiding extra artificial gauge choice singularities in additional to those actual Wilson loop interpolation singularities explained in Section \ref{ss_interpolation}.

Let us elaborate on this point. In \cite{Luscher:1981zq}, for each given hypercube $h$, the $u_l$ on the links are put into the ``complete axial gauge associated with $h$'' to begin with.
\footnote{For the same link $l$, when computing the instanton density over different hypercubes that contain $l$, the $u_l$ is therefore put into different gauges.}
Then, by specifying the Wilson lines $P$ and $S$, gauge choices in the interior of the plaquettes and cubes are essentially being made---in particular, given a Wilson line $w$ of length $1$ running in $\mu$-direction, a Wilson line running along the first $y \ (y<1)$ portion of it (see any segment of $S$ in Fig.~\ref{fig:S} or of $P$ in Fig.~\ref{fig:P_def}) is chosen to be $w^y$. However, there would then be artificial singularities not due to any Wilson loop interpolation (those explained in Section \ref{ss_interpolation}) but merely due to the gauge choice---for the particular $w^y$ gauge choice, the gauge choice singularity develops when $w\rightarrow -1$. The actual Wilson loop interpolation singularity and the artificial gauge choice singularity are not distinguished in \cite{Luscher:1981zq}, but the gauge choice singularities can actually be avoided if we have an expression that is manifestly independent of the gauge choice. On physical grounds, in our construction we only demand CS sensitivity $|\nu|\rightarrow 0$ when actual Wilson loop interpolation singularity occurs.
\end{itemize}

\begin{figure}
    \centering
    \includegraphics[width=0.5\linewidth]{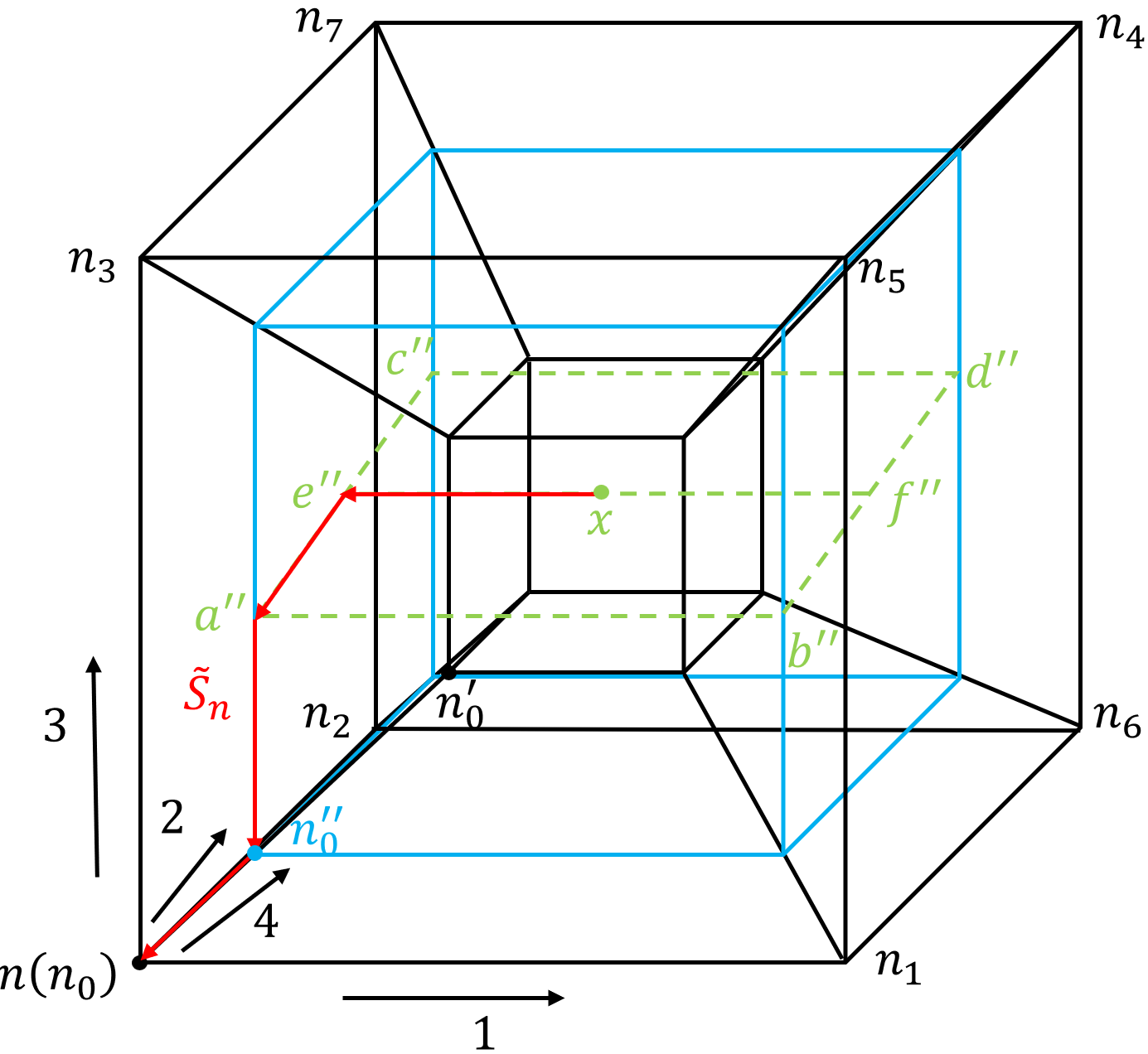}
    \caption{Definition of $\wt{S}_{n}(x)$. Here we are again picturing a 4d hypercube, with the vertices labeled the same as Fig.~\ref{fig:T}. But unlike in Fig.~\ref{fig:T}, now $x$ can be any point in the hypercube $h(n)$ instead of being restricted in $c(n,1)$. The red Wilson line $\wt{S}_{n}(x)$ from $x$ to $n_0$ consists of four segments that go backwards along the $\mu=1, 2, 3, 4$ directions consecutively.}
    \label{fig:S_h}
\end{figure}

Given these motivations, we separate out the gauge dependence in (\ref{eqn:Luscher_q_density}) via the following. Given the gauge field configuration on $\partial h(n)$, we can always find some smooth interpolation of the gauge field into the interior of $h(n)$ \cite{Dijkgraaf:1989pz}.  
\footnote{We are not saying there is a single protocol which gives a smooth interpolation into $h$ such that the interpolation always varies smoothly as we vary the field configuration on $\partial h(n)$. We are given a fixed (instead of varying) configuration on $\partial h$ and asking for the existence of a interpolation into $h$ that is smooth in the interior of $h$.}
We only need the conceptual existence of such a continuous gauge field interpolation into the hypercube to perform the derivation, but we will not need any particular construction of it. Given such an interpolation, for any $x\in h(n)$, we define $\wt{S}_{n}(x)$ as in Fig.~\ref{fig:S_h}, i.e.
\begin{equation}
    \wt{S}_{n}(x)\equiv (00''a''e''x) .
\end{equation}
Then we compare $\wt{S}_{n}(x)$ to $S_{n,\mu}$ when $x$ lands on any cube on the boundary of $h(n)$.\footnote{We have implicitly required that $\wt{S}_{n}(x)$ is compatible with $S_{n,\mu}(x)$ when the Wilson line entirely lies on $\partial h(n)$. This condition fixes the value of $\wt{S}_{n}(x)$ for $x$ on $\partial h(n)$ except for on $c(n+\hat 1,1)$, where the interpolation can be freely chosen so it is always possible to find one under this compatibility condition.}
In $c(n,\mu) \ (\mu=1,2,3,4)$ we have
\begin{equation}
    \wt{S}_{n}(x)=S_{n,\mu}(x).
\end{equation}
In $c(n+\hat 4,4)$ we have
\begin{equation}
    \wt{S}_{n}(x)=(00')S_{n+\hat 4,4}(x).
\end{equation}
In $c(n+\hat 3,3)$ we have
\begin{equation}
    \wt{S}_{n}(x)=(00''3''3)S_{n+\hat 3,3}(x)=R_{n,1;3}(x-\hat{1})S_{n+\hat 3,3}(x)=R_{n,2;3}(x-\hat{2})S_{n+\hat 3,3}(x)
\end{equation}
In $c(n+\hat 2,2)$ we have
\begin{equation}
    \wt{S}_{n}(x)=(00''a''c''2''2)S_{n+\hat 2,2}(x)=R_{n,1;2}(x-\hat{1})S_{n+\hat 2,2}(x).
\end{equation}
Finally in $c(n+\hat 1,1)$ we have
\begin{equation}
    \wt{S}_{n}(x)= T_n(x)S_{n+\hat 2,2}(x)
\end{equation}
where $T_n(x)\equiv (00''a''e''f''b''1''1)$ has been introduced in the main text.

Substituting $S$ in (\ref{eqn:Luscher_q_density}) with $\wt{S}$ by the equations above, and using the identity
\begin{equation}
\begin{aligned}
    &\epsilon^{\mu\nu\rho\sigma}\Tr\left[(vw)^{-1}\partial_\nu (vw)(vw)^{-1}\partial_\rho (vw)(vw)^{-1}\partial_\sigma (vw)\right]\\
    &\quad =\epsilon^{\mu\nu\rho\sigma}\Tr\left\{(w^{-1}\partial_\nu w)(w^{-1}\partial_\rho w)(w^{-1}\partial_\sigma w)\right.\\
    & \qquad \left.+(v^{-1}\partial_\nu v)(v^{-1}\partial_\rho v)(v^{-1}\partial_\sigma v)-3\partial_\nu\left[(v^{-1}\partial_\rho v)(\partial_\sigma w \, w^{-1})\right]\right\}
\end{aligned}
\end{equation}
to carefully cancel the integrals on plaquettes, we find
\begin{equation}
\begin{aligned}
    &\hspace{.9cm} (\mathrm{d}\alpha^{(0)})_{h(n)} \\
    & = \ \ \frac{1}{12\pi}\int_{\partial h(n)}\Tr\left[\wt{S}_{n}^{-1}\mathrm{d}\wt{S}_{n}\wedge \wt{S}_{n}^{-1}\mathrm{d}\wt{S}_{n}\wedge \wt{S}_{n}^{-1}\mathrm{d}\wt{S}_{n}\right]\\
    &\qquad -\frac{\epsilon^{1\mu\nu\rho}}{12\pi}\int_{c(n+\hat 1,1)}\mathrm{d}^3 x \: \Tr\left[T_n(x)^{-1}\partial_\mu T_n(x) T_n(x)^{-1}\partial_\nu T_n(x) T_n(x)^{-1}\partial_\rho T_n(x)\right]\\
    &\qquad -\frac{\epsilon^{1\mu\nu\rho}}{4\pi}\int_{p(n+\hat 1+\hat \mu,1,\mu)}\mathrm{d}^2 x \: \Tr\left[ T_n(x)^{-1} \left( R_{n,1;\mu}(x-\hat{1})  \partial_\nu R_{n+\hat \mu,\mu;1}(x) \partial_\rho R_{n+\hat 1,1;\mu}(x)^{-1} \right. \right. \\
    & \phantom{\int} \left.\left.  \hspace{6.4cm} + \partial_\nu R_{n,1;\mu}(x-\hat{1})  R_{n+\hat \mu,\mu;1}(x) \partial_\rho R_{n+\hat 1,1;\mu}(x)^{-1} \right. \right. \\
    & \phantom{\int} \left.\left. \hspace{6.4cm} + \partial_\nu R_{n,1;\mu}(x-\hat{1}) \partial_\rho R_{n+\hat \mu,\mu;1}(x) R_{n+\hat 1,1;\mu}(x)^{-1} \right) \right] \ . \\
\end{aligned}
\label{eqn:q_density_app}
\end{equation}
Now we make two key observations:
\begin{itemize}
\item Except for the first term, the remaining terms only involve Wilson lines that start and end at lattice vertices, and it is therefore easy to see these  terms are gauge invariant, because those gauge transformation made at the vertices are not acted on by the derivatives (as in Eq.(\ref{eqn:R_gauge_transf})). Therefore, all the gauge dependence is in the first term. 
\item The first term is a WZW integral over a closed 3d manifold, which gives $2\pi\mathbb{Z}$. Note for this to be a valid statement, we need to ensure $\wt{S}_{n}(x)$ is \emph{smoothly defined} over $x\in \partial h(n)$---and it is indeed so, because: 1) as we argued before introducing $\wt{S}$, there exists a smooth interpolation of the gauge field into the interior of $h$, 2) when $x$ varies in $h(n)$ (not just in $\partial h(n)$), the chosen path that the Wilson line $\wt{S}$ goes along varies smoothly, and 3) it is always possible to make a smooth gauge choice (even though we do not need to specify the choice) for any given gauge field configuration in the hypercube because a hypercube is contractible. Furthermore, in our case since $\wt{S}_{n}$ is smoothly interpolated into the hypercube, this WZW integral simply gives zero.
\end{itemize}
Putting these two observations together, we arrive at the manifestly gauge invariant Eq.(\ref{eqn:q_density}).

\

To prove the crucial property Eq.(\ref{eqn:sum_q_int}), we shall go back to Eq.(\ref{eqn:Luscher_q_density}). We can see that according to Eq.(\ref{eqn:Luscher_q_density}),
\begin{equation}
    (\mathrm{d}\alpha^{(0)})_{h}=\sum_{c\in \partial h}X_c = dX_h
    \label{eqn:q_dX}
\end{equation}
where for $c=c(m,\mu)$, $X_c$ is defined as (here we actually implicitly took the orientation into account, so that $X_{-c}=-X_c$.)
\begin{equation}
\begin{aligned}
    X_c=&\sum_{\nu,\rho,\sigma}\frac{\epsilon^{\mu\nu\rho\sigma}}{12\pi} \left\{\int_{c(m,\mu)}\mathrm{d}^3x \: \Tr[(S_{m,\mu})^{-1}\partial_\nu S_{m,\mu}(S_{m,\mu})^{-1}\partial_\rho S_{m,\mu}(S_{m,\mu})^{-1}\partial_\sigma S_{m,\mu}]\right.\\
    &\left.+3\int_{p(m+\hat{\nu},\mu,\nu)}\mathrm{d}^2x \: \Tr[P_{m+\hat{\nu},\mu,\nu}\partial_\rho(P_{m+\hat{\nu},\mu,\nu})^{-1} (R_{m,\mu;\nu})^{-1}\partial_\sigma R_{m,\mu;\nu}]\right\}.
\end{aligned}
\end{equation}
Thus we indeed have
\begin{equation}
    \sum_h (\mathrm{d}\alpha^{(0)})_{h(n)} =\sum_h dX_h = 0.
\end{equation}
(In \cite{Luscher:1981zq}, the Wilson line $S$ on a cube $c$ is calculated under different gauges when $c$ appears on the boundary of its two neighboring hypercubes,
\footnote{More precisely, in \cite{Luscher:1981zq}, given the gauge choice of the link variables $u_l$ around a hypercube, the protocol to determine the gauge choice in the interior of the cubes around that hypercube is fixed and independent of the hypercube; however, the gauge choice of $u_l$ itself indeed depends on the hypercube to begin with: When calculating $(\mathrm{d}\alpha^{(0)})_{h}$ using (\ref{eqn:Luscher_q_density}), $u_l$ will take the ``complete axial gauge associated with $h$'', hence for a same link $l$, when viewed from different $h$, $u_l$ will be under different gauges, hence so will $S$.}
hence the cancellation is not exact, but might leave a $2\pi\mathbb{Z}$ part. But here we are only using the $\mod 2\pi\mathbb{Z}$ part anyways.) Note that $X_c \mod 2\pi$ roughly plays the role of the CS phase saddle $\alpha^{(0)}_c$. We will discuss more about the CS phase saddle $\alpha^{(0)}_c$ in the next appendix.

\section{Construction of the CS Phase Saddle $e^{i\alpha_c^{(0)}}$ on a Single Cube}
\label{app:CS_cube}

To construct the CS phase saddle $\alpha^{(0)}_c\equiv\arg(\nu_c)$ on a cube $c$, it turns out convenient to use the following ``fictitious hypercube'' procedure.

\begin{figure}
    \centering
    \includegraphics[width=0.35\linewidth]{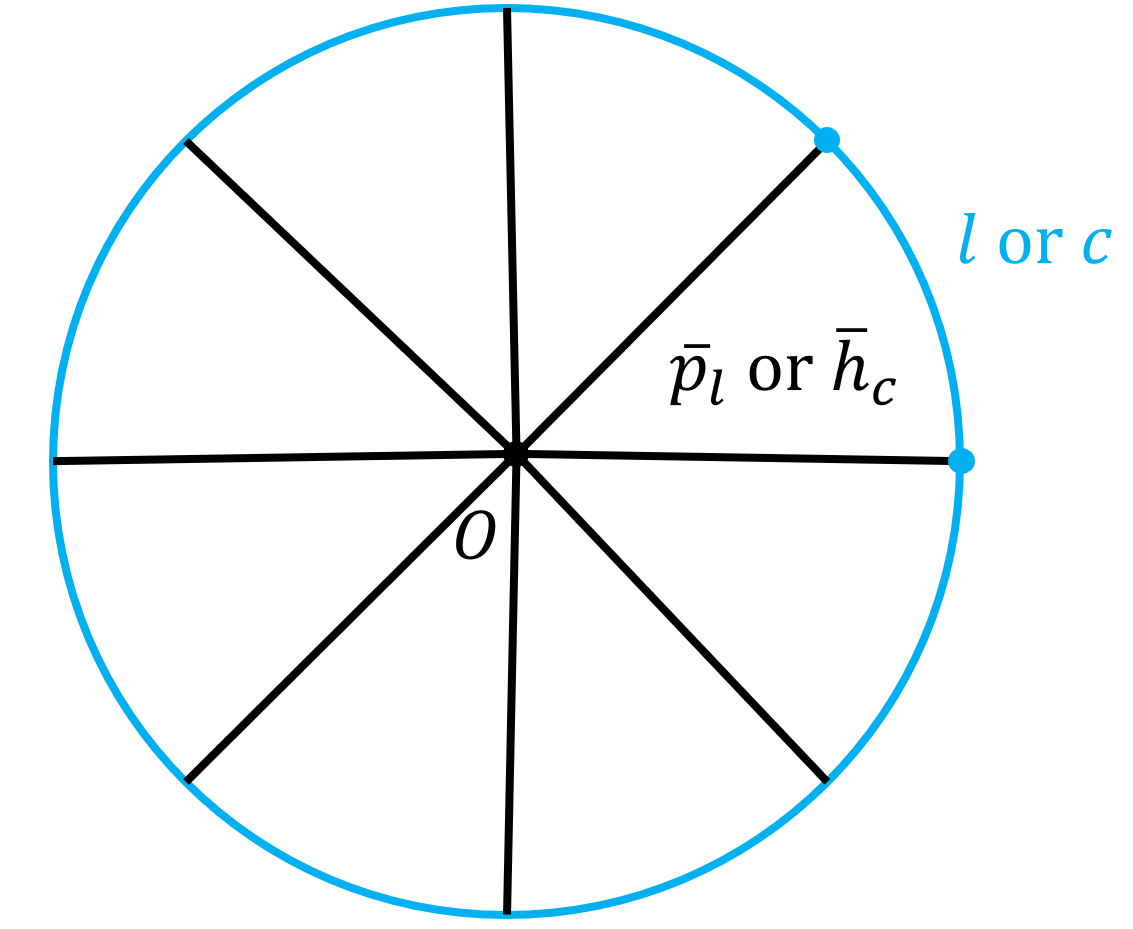}
    \caption{A sketch of how to construct the CS phase saddle on a cube $c$ using fictitious hypercube $\bar h_c$. The idea is the same as representing the connection on link $l$ for a $U(1)$ lattice gauge theory using the flux on fictitious plaquette $\bar p_l$. The blue links represent the physical lattice, while the black links are fictitious.}
    \label{fig:fictitious_sketch}
\end{figure}

Let us motivate the procedure by the simpler case of $U(1)$ gauge field in 1d. Suppose we have a 1d loop in the continuum, discretized into many segments, see the blue circle in Fig.~\ref{fig:fictitious_sketch}. We can describe the Wilson line along a segment $l$ as $u_l=e^{i\int_l A}$. Alternatively, we can introduce a fictitious point $O$ in a higher dimension and connect it with the physical vertices via fictitious links, see the black lines in Fig.~\ref{fig:fictitious_sketch}. The Wilson lines on the fictitious links are fixed. Then we may as well define $u_l=e^{i\int_{\bar{p}_l} F}=e^{i\int_{\partial \bar p_l}A}$, the $U(1)$ flux over the ``pizza slice'' $\bar{p}_l$ made of a physical link $l$ and two fictitious links. $u_l$ should conceptually be gauge dependent, while $F$ is gauge invariant, but there is no contradiction here---the gauge dependence of $u_l$ becomes the dependence of $F$ on our choice of the gauge field on the fictitious links. Also it is straightforward to see that this gauge dependence is canceled if we calculate the holonomy around the entire physical loop, $\prod_l u_l$, since adjacent ``pizza slices'' share the same fictitious link.

\begin{figure}
    \centering
    \includegraphics[width=0.4\linewidth]{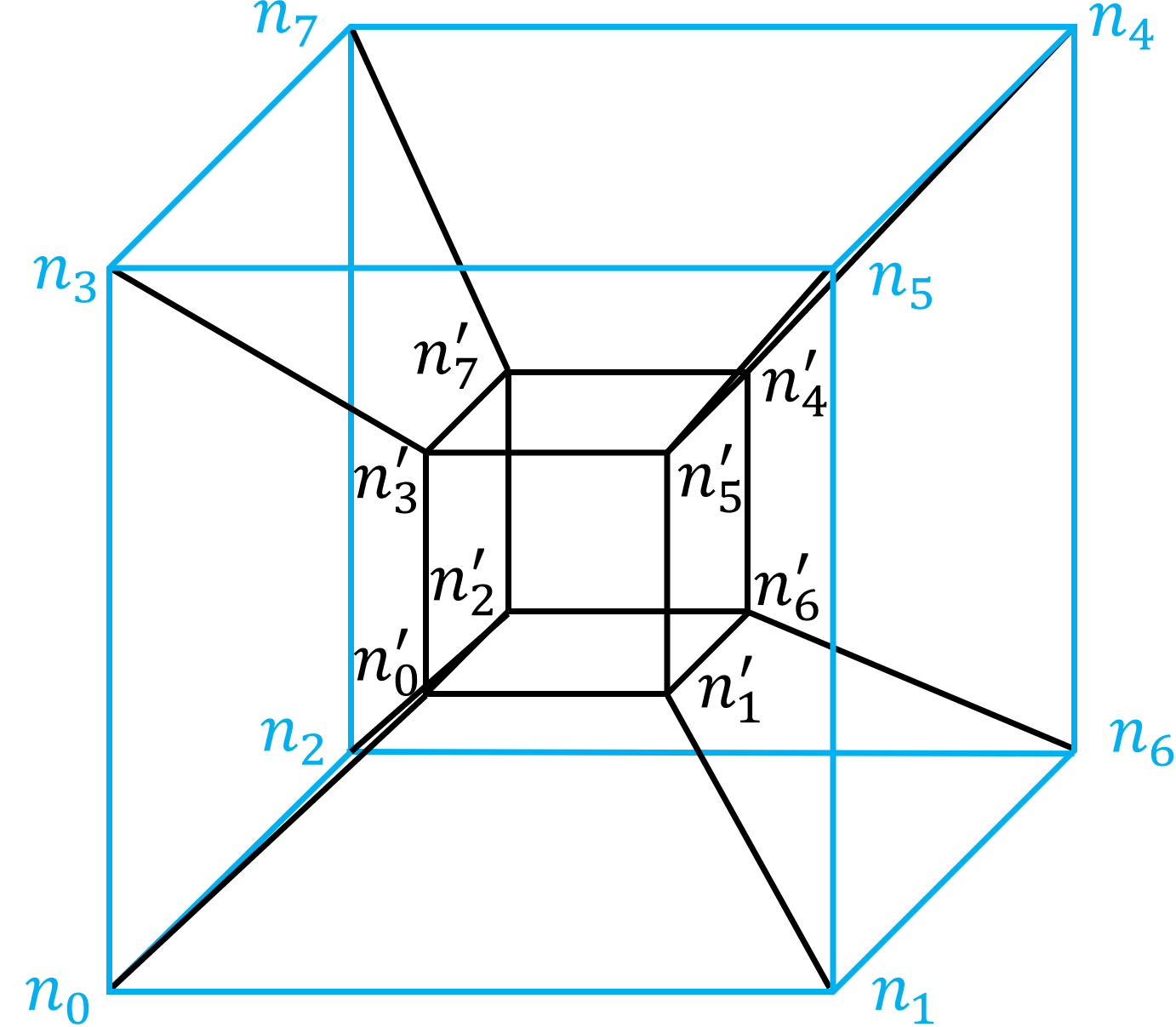}
    \caption{The fictitious hypercube $\bar h_c$. The blue part is the physical cube $c$ and the black links are the fictitious links.}
    \label{fig:fictitious_hypercube}
\end{figure}

Similarly, for an $SU(2)$ gauge theory, to construct the CS phase saddle $\alpha^{(0)}_c$ in a cube $c$, we connect the physical vertices to a fictitious point $O$ to form a fictitious hyperpyramid, and then we can divide the hyperpyramid into two parts: one hyperpyramid with trivial connection all over, and then one hypercube $\bar h_c$ which contains the physical cube as well as some fictitious cubes, see Fig.~\ref{fig:fictitious_hypercube}. Then we set
\begin{equation}
    e^{i\alpha^{(0)}_c} \equiv e^{i(\mathrm{d}\alpha^{(0)})_{\bar h_c}}.
\end{equation}
This means that we can construct the CS phase saddle around the fictitious hypercube $\bar{h}_c$ to represent the CS phase saddle on the physical cube $c$.

Analogous to the $U(1)$ gauge theory case, here the CS phase saddle should conceptually be $SU(2)$ gauge dependent, and the dependence is manifested as the dependence of $e^{i(\mathrm{d}\alpha^{(0)})_{\bar h_c}}$ on the choice of fictitious fields on those fictitious links and fictitious plaquettes. The gauge dependence will cancel if we take product of the CS phase saddle over a closed 3d surface, $\prod_c e^{i\alpha^{(0)}_c}=\prod_c e^{i\mathrm{d}\alpha^{(0)}_{\bar h_c}}$,
\footnote{The closed 3d surface can be a contractible one, such as the boundary of a physical 4d hypercube as well as a non-contractible one.}
because by Eq.(\ref{eqn:q_dX}), the contributions from the $X_{\bar{c}_p}$ on fictitious cubes cancel out. This is crucially important, because this means the $SU(2)$ gauge dependence of $e^{i \alpha^{(0)}_c}$ is realized as a 2-form $U(1)$ gauge transformation that can  henceforth be absorbed by $e^{i\alpha_c}$ in (\ref{eqn:W3}):
\begin{equation}
e^{i\alpha^{(0)}_c} \rightarrow e^{i\alpha^{(0)}_c+i\mathrm{d}\zeta_c}, \ \ \ \ e^{i\alpha_c} \rightarrow e^{i\alpha_c+i\mathrm{d}\zeta_c}, \ \ \ \ e^{i\zeta_p}\in U(1) \ .
\label{eqn:alpha_gauge}
\end{equation}
This keeps the total CS phase $\prod_c e^{i\alpha_c}$ over an closed 3d lattice and the physical instanton density (\ref{eqn:q_latt}) on a physical 4d hypercube invariant. $e^{i\zeta_p}$ can be interpreted as the change of $X_{\bar{c}_p}$ on the fictitious cubes when we change the choice of fictitious fields around $\bar{c}_p$.

Let us discuss more about the fictitious gauge field on the fictituous cells. For those ``purely fictitious links'' $\tilde{l}$ which do not touch any physical cell, such as the link $\langle 0'1'\rangle$ in Fig.~\ref{fig:fictitious_hypercube}, we can simply set $u_{\tilde{l}}=1$, and for those ``purely fictitious plaquette'' $\tilde{p}$ which do not touch any physical cell, such as the plaquette $(0'3'5'1'0')$ in Fig.~\ref{fig:fictitious_hypercube}, we simply set $\xi_{\tilde{p}}=+$.

Those ``partially fictitious links'' $\bar l_n$ and ``partially fictitious plaquettes'' $\bar p_l$ which touch the physical cells are more interesting. We may as well fix the link fields to be $u_{\bar{l}_n}=1$ and plaquette fields to be $\xi_{\bar{p}_l}=+$ as above, but now  $u_{\bar{p}_l}=u_l$ might turn out to be $-1$, and together with $\xi_{\bar{p}_l}=+$ the fictitious plaquette interpolation runs into singularity, not rescued by any weight $W_2\rightarrow 0$ since there is just no weight on the fictitious plaquette. Even if the fictitious plaquette interpolation does not run into singularity, the fictituous cube interpolation in $\bar{c}_p$ may still run into those singularities mentioned in Section \ref{ss_interpolation}, but now not rescued by any $|\nu|\rightarrow 0$ as in Section \ref{ss_CS_sensitivity}, because again there is just no weight on the fictitious cube. In fact, these singularities are nothing but reminiscence of those gauge choice singularities that we tried to get rid of in the previous appendix.

At a conceptual level, these gauge choice singularities do not matter, because any physical observable will be gauge invariant, and these gauge choice singularities will be ``washed away'' by the fluctuation of $e^{i\alpha_c}$, just like the apparent singularity in the logarithm in $A_l=-i\ln u_l$ in the Villainized $U(1)$ gauge theory will be ``washed away'' by the fluctuation of $s_p$ in $F_p=(\mathrm{d}A)_p+2\pi s_p$ \cite{Chen:2024ddr}. At the numerical implementation level, gauge choice singularities might lead to practical problems, therefore in 4d Yang-Mills application we prefer to directly construct the manifestly gauge invariant $e^{i \mathrm{d}\alpha^{(0)}_h}$, rather than $e^{i \alpha^{(0)}_c}$. On ther other hand, for 3d Chern-Simons-Yang-Mills we can only use $e^{i \alpha^{(0)}_c}$, but the CS theory is not going to be calculated in Monte-Carlo anyways due to its complex phase, so there is no numerical practical consideration to begin with.

In fact, if we really want to, we can still get rid of these gauge choice singularities in the construction of $e^{i \alpha^{(0)}_c}$. One can introduce some additional weights in the partition function to average over different gauge choices so that there is manifestly no singularity given any physical gauge field configuration. Since this further refinement is quite tedious, and we are not going to really use it in practice, we will not elaborate on it here.

\bibliography{LattQCDIns_explicit}{}
\bibliographystyle{JHEP}

\end{document}